\documentclass[10pt, conference]{IEEEtran}
\IEEEoverridecommandlockouts

\usepackage{cite}
\usepackage{amsmath,amssymb,amsfonts}
\usepackage{amsthm}
\usepackage{algorithmic}
\usepackage{textcomp}
\usepackage{xcolor}
\usepackage{graphics}
\usepackage{textcomp}
\usepackage{comment}
\usepackage{tabularx}
\usepackage{breqn}
\usepackage{url}
\usepackage{listings}
\usepackage{threeparttable}
\usepackage[linesnumbered,ruled]{algorithm2e}
\usepackage{color}
\usepackage{soul}
\usepackage{multirow}

\usepackage{tikz} \usepackage{lipsum} \makeatletter \def\ieeecopyright{ \footnotesize © 2023 IEEE. Personal use of this material is permitted.\newline DOI: 10.1109/ISORC58943.2023.00015} \makeatother \AddToHook{shipout/firstpage}{\begin{tikzpicture}[remember picture,overlay] \node[anchor=south west,xshift=1.0cm,yshift=0.8cm] at (current page.south west){\parbox{\linewidth}{\raggedright\ieeecopyright}}; \end{tikzpicture} }

\renewcommand{\baselinestretch}{0.985}
\makeatletter
\newcommand{\figcaption}[1]{\def\@captype{figure}\caption{#1}}
\newcommand{\tblcaption}[1]{\def\@captype{table}\caption{#1}}
\makeatother

\def\BibTeX{{\rm B\kern-.05em{\sc i\kern-.025em b}\kern-.08em
    T\kern-.1667em\lower.7ex\hbox{E}\kern-.125emX}}

\sethlcolor{yellow}

\newcommand{\rr}[2]{#2}

\newcommand{\dq}[1]{``#1''}

\newcommand{\la}[0]{$\leftarrow$ }
\newcommand{\ch}[0]{\checkmark}

\newcommand{\tabml}[1]{\hspace{-2.2mm}\begin{tabular}{l} #1 \end{tabular}}

\newcommand{\csix}[1]{#1}

\def\MR#1#2{\multirow{#1}{*}{#2}}
\def\MC#1#2#3{\multicolumn{#1}{#2}{#3}}

\begin{document}

\title{RD-Gen: Random DAG Generator \\ Considering Multi-rate Applications for \\ Reproducible Scheduling Evaluation}

\author{
    \IEEEauthorblockN{Atsushi Yano}
    \IEEEauthorblockA{\textit{Tier IV, Inc}\\
        atsushi.yano.2@tier4.jp}
    \and
    \IEEEauthorblockN{Takuya Azumi}
    \IEEEauthorblockA{\textit{Tier IV, Inc}\\
        \textit{Graduate School of Science and Engineering,} \\
        \textit{Saitama University}}
    \thanks{This paper is based on results obtained from a project subsidized by the New Energy and Industrial Technology Development Organization (NEDO).}
}

\maketitle

\begin{abstract}
    Real-time systems have various requirements such as the deadline and resource constraints.
    In addition, real-time systems are becoming larger and more complex, and studies on performance analysis and efficient scheduling algorithms are becoming increasingly important.
    Directed acyclic graph (DAG) models, which can express task dependencies and parallelism, are used for such studies.
    Random DAG sets are used to demonstrate the effectiveness and objectivity of methods proposed for real-time systems.
    However, there is no random DAG generation tool available that can generate a DAG set that considers the latest multi-rate applications.
    Therefore, researchers need to generate random DAG sets on their own, leading to additional effort and reduced reliability and reproducibility.
    To solve this problem, we propose a random DAG generator considering multi-rate applications for reproducible scheduling evaluation (RD-Gen).
    RD-Gen also enables batch generation of random DAG sets with different parameters.
    Case studies are used to demonstrate that RD-Gen can manage various problem settings and DAG study requirements.
\end{abstract}


\vspace{-2mm}
\section{Introduction}
\label{sec: introduction}
\vspace{-1mm}

Real-time systems, such as self-driving systems, need to successfully execute while meeting various requirements such as producing an output within a pre-determined time (i.e., meeting deadlines), having low power consumption, and meeting resource constraints~\cite{senapati2021hmds}.
To fulfill these constraints, many studies have been conducted on task allocation and scheduling, as well as on analyzing the end-to-end latency and the response time of a system~\cite{kordon2020evaluation}.
Systems are increasingly becoming larger and more complex, and many studies use models, such as directed acyclic graphs (DAGs), to represent the complex dependencies and parallelism of the tasks in these systems.

DAGs are used in many allocations, scheduling, and latency analysis studies~\cite{choi2021picas, klaus2021constrained} because they express the process flow from system input to output and can represent various types of information such as the dependencies between tasks, the task execution time, and the period.
To evaluate the performance of proposed methods, it is important to compare them with existing methods using task sets.
Accordingly, in method evaluations using DAGs, randomly generated DAGs are used to ensure objectivity and demonstrate generality~\cite{he2021response, verucchi2020latency, senapati2021hmds}.

To aid in such evaluations, random DAG generation tools, such as task graph for free (TGFF)~\cite{tgff} and GGen~\cite{cordeiro2010random}, have been proposed and utilized in the latest publications~\cite{sun2021deepweave, huang2020hda}.
These tools allow the user to parametrically specify the shape of a DAG and the properties assigned to the tasks and edges.
% Furthermore, because these tools use a pseudo-random number generator, other researchers can easily reproduce a given DAG set by specifying the same options.
However, TGFF and GGen were proposed in 1999 and 2010, respectively, and cannot meet the requirements of multi-rate DAGs considering state-of-the-art real-time systems.

Because embedded systems in automobiles and avionics, as well as in self-driving systems, contain multiple tasks that operate over different periods (e.g., localization~\cite{verucchi2020latency}), studies targeting multi-rate DAGs are becoming increasingly important~\cite{gunzel2021suspension, kordon2020evaluation}.
In studies of such multi-rate DAGs, not only the shape of the DAG but also the ratio of the execution time to the task period has a significant impact on the performance of the method (e.g., implicit deadlines~\cite{ueter2021hard, cho2021conditionally} and task utilization~\cite{yang2020mixed}).
However, TGFF cannot generate multi-rate DAGs, and GGen can only randomly set the period and the execution time of tasks.
Therefore, most researchers who consider multi-rate DAGs have to implement their own random DAG sets~\cite{voronov2021ai, dong2019efficient, yang2020mixed}.
It is laborious for researchers to prepare their own random DAG sets, and this practice further reduces the reliability and reproducibility of the evaluation results.

To solve these problems, this paper proposes a random DAG generator considering multi-rate applications for reproducible scheduling evaluation (RD-Gen).
RD-Gen extends existing DAG generation methods and provides a flexible evaluation platform.
Because RD-Gen uses a pseudo-random number generator, other researchers can reproduce the DAG sets used in an evaluation by specifying the same options.

\textbf{\rr{1.1, 2.1, 3.1}{Contributions:} } Our primary contributions are summarized as follows.
\vspace{-1mm}
\begin{itemize}
    \item RD-Gen extends existing random DAG construction methods, {\it Fan-in/Fan-out}~\cite{tgff} and {\it G(n, p)}~\cite{cordeiro2010random} methods, to meet researchers' requirements.
    \item RD-Gen proposes a new {\it Chain-based} method to flexibly construct state-of-the-art chain-based multi-rate DAGs.
    \item RD-Gen reduces implementation effort through the automatic setting of complex parameters and batch generation of random DAG sets.
\end{itemize}
\vspace{-1mm}

The remainder of the paper is organized as follows.
Section~\ref{sec: system_model} describes a system model.
Section~\ref{sec: design_implementation} explains the design and implementation of RD-Gen.
Section~\ref{sec: case_study} presents case studies.
Section~\ref{sec: evaluation} compares RD-Gen with existing random DAG generation methods.
Section~\ref{sec: related_work} discusses related work.
Finally, Section~\ref{sec: conclusion} presents the conclusions and future work.

\vspace{-1mm}
\section{System model}
\label{sec: system_model}
\vspace{-1mm}

\begin{figure}[tb]
    \centering
    \scalebox{1.0}{
        \includegraphics[width=\linewidth]{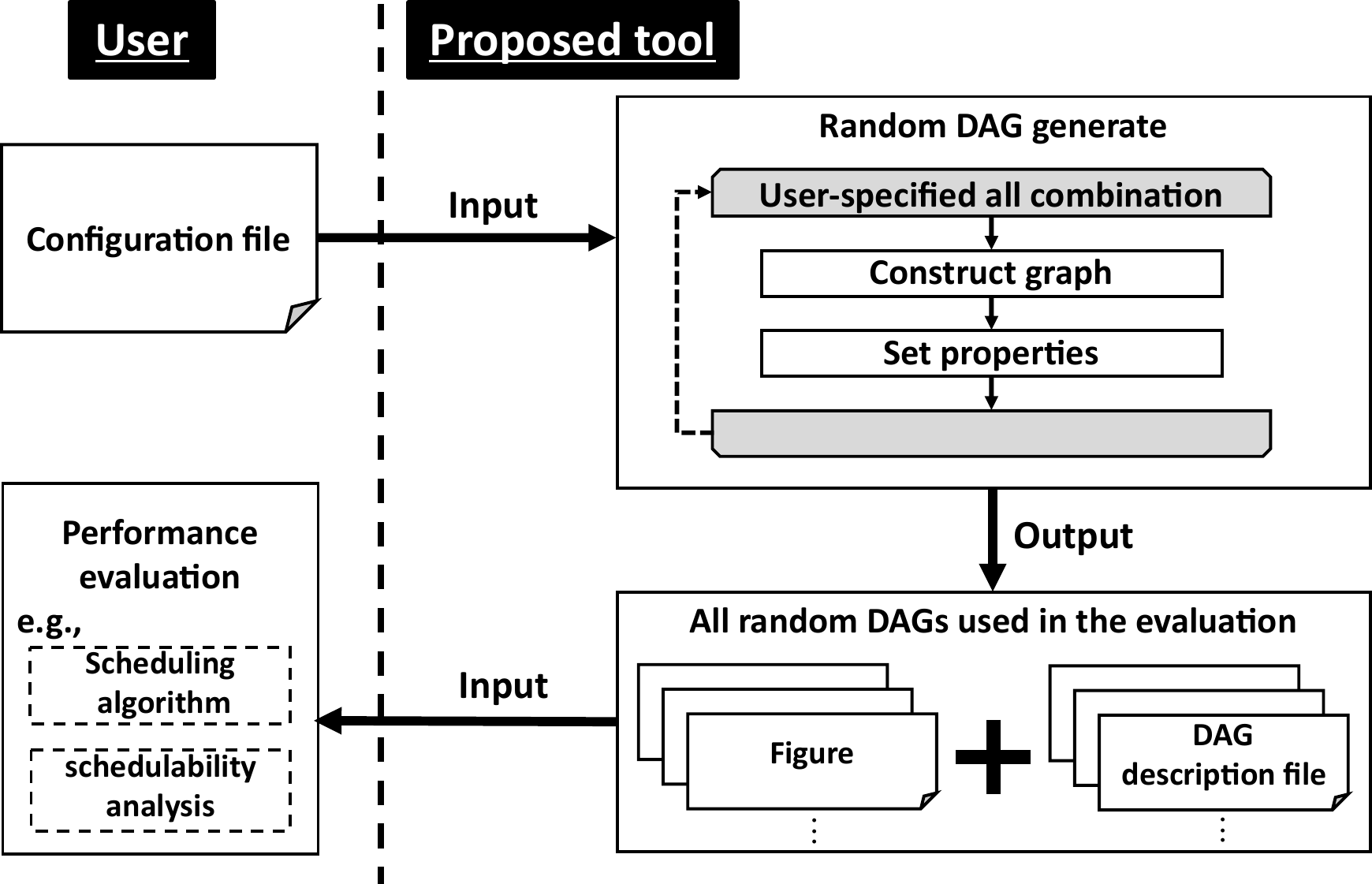}
    }
    % \vspace{-5.5mm}
    \caption{System model.}
    \label{fig: system_model}
    % \vspace{-4mm}
\end{figure}

\begin{table}[tb]
    \centering
    \caption{DAG Notations}
    \label{tab: dag_notations}
    \vspace{-3mm}
    \renewcommand{\arraystretch}{1.25}
    \scalebox{0.83}{
        \begin{tabular}{|c|c|l|}\hline
            \textbf{DAG Types}   & \textbf{Symbols} & \MC{1}{c|}{\textbf{Descriptions}}                  \\\hline
            \MR{10}{Common}      & $G$              & DAG                                                \\ \cline{2-3}
                                 & $V$              & Set of all nodes of $G$                            \\ \cline{2-3}
                                 & $|V|$            & Total number of nodes of $G$                       \\ \cline{2-3}
                                 & $E$              & Set of all edges of $G$                            \\ \cline{2-3}
                                 & $v_i$            & $i$-th node                                        \\ \cline{2-3}
                                 & $C_i$            & Worst-case execution time (WCET) of $v_i$          \\ \cline{2-3}
                                 & $e_{i,j}$        & Edge from $v_i$ to $v_j$                           \\ \cline{2-3}
                                 & $comm_{i,j}$     & Communication time of $e_{i,j}$                    \\\cline{2-3}
                                 & $CCR$            & Communication-to-computation ratio (CCR) of $G$    \\\cline{2-3}
                                 & $D(G)$           & End-to-end deadline                                \\ \hline
            \MR{5}{Multi, Chain} & $v^{td}_i$       & $i$-th timer-driven node                           \\ \cline{2-3}
                                 & $\phi_i$         & Offset of $v^{td}_i$                               \\ \cline{2-3}
                                 & $T_i$            & Period of $v^{td}_i$                               \\ \cline{2-3}
                                 & $u_i$            & Utilization of $v^{td}_i$                          \\\cline{2-3}
                                 & $U$              & Total utilization of $G$                           \\\hline
            \MR{6}{Chain}        & $v^{ed}_i$       & $i$-th event-driven node                           \\ \cline{2-3}
                                 & $\Gamma_i$       & $i$-th chain                                       \\\cline{2-3}
                                 & $|\Gamma|$       & Total number of chains of $G$                      \\\cline{2-3}
                                 & $C_{\Gamma_i}$   & WCET of $\Gamma_i$                                 \\ \cline{2-3}
                                 & $u_{\Gamma_i}$   & Utilization of $\Gamma_i$                          \\ \cline{2-3}
                                 & $T_{\Gamma_i}$   & Period of the head timer-driven node of $\Gamma_i$ \\ \hline
        \end{tabular}
    }
    \vspace{-6mm}
\end{table}

This section describes the system model of this paper.
The overview of this paper is shown in Fig.~\ref{fig: system_model}.
First, Section~\ref{ssec: single_rate_dag} describes the basic single-rate DAG.
Next, Section~\ref{ssec: multi_rate_dag} presents a multi-rate DAG in which all nodes are timer-driven nodes.
Finally, Section~\ref{ssec: chain_based_dag} explains multi-rate DAGs consisting of timer-driven and event-driven nodes found in self-driving systems.
For clarity, a multi-rate DAG consisting of only timer-driven nodes is simply called a multi-rate DAG, while a DAG consisting of a chain of timer-driven and event-driven nodes is called a chain-based DAG hereafter.
The DAG notations in this paper are listed in Table~\ref{tab: dag_notations}.

\vspace{-1mm}
\subsection{Single-rate DAG}
\label{ssec: single_rate_dag}
\vspace{-1mm}

Single-rate DAGs are DAGs with either a single source node or, all source nodes entering at the same time.
% Here, the source node represents the input to the system (e.g., a sensor event or a command from the user), and the sink node represents the final output of the system.

A DAG consists of a node set and an edge set, denoted $G = (V, E)$.
Nodes represent tasks in the system, and edges represent communication and dependencies between nodes and \rr{3.5}{precedence constraints}.
$V$ is the set of all nodes, expressed as $V = \{v_1, ..., v_{|V|}\}$, where $|V|$ is the total number of nodes.
Each node has a worst-case execution time (WCET), and the WCET of $v_i$ is denoted as $C_i$.
$E$ is the set of all edges, where each edge $e_{i,j} \in E$ represents communication between $v_i$ and $v_j$ and a \rr{3.5}{precedence constraint}.
When $e_{i,j}$ exists in the DAG, $v_j$ cannot be executed until $v_i$ has completed its execution and the output of $v_i$ has arrived.
If the communication time is given as an assumption, the communication time at $e_{i,j}$ is denoted as $comm_{i, j}$.
The ratio of the sum of the communication times of all edges to the sum of the execution times of all nodes is called the communication-to-computation ratio (CCR) and is defined by Eq.~(\ref{eq: ccr}).

\vspace{-8mm}
\begin{equation}
    \label{eq: ccr}
    CCR = \frac{\sum\limits_{e_{i,j} \in E}comm_{i, j}}{\sum\limits_{v_i \in V}C_i}
\end{equation}
\vspace{-2mm}

An end-to-end deadline $D(G)$ is set at the sink node when the safety of hard real-time systems~\cite{yano2021work} or the cloud computing the quality of service~\cite{zhang2020efficient} must be guaranteed.

\vspace{-1mm}
\subsection{Multi-rate DAG}
\label{ssec: multi_rate_dag}
\vspace{-1mm}

Multi-rate DAGs are DAGs consisting of timer-driven nodes that operate at different periods.
In multi-rate DAGs, while the basic definitions of node and edge are the same as in those given in Section~\ref{ssec: single_rate_dag}, the meaning of edge is different.
Edges indicate communication between nodes using shared memory, where each node reads data at the start of execution and writes data at the end of execution~\cite{klaus2021constrained, verucchi2020latency, kordon2020evaluation}.
Although undersampling and oversampling occur due to differences in periods, the latest data are used according to {\it last-is-best} semantics~\cite{klaus2021constrained}.

Each timer-driven node in such multi-rate DAGs is denoted by $v^{td}_i$, and $v^{td}_i$ is characterized by the tuple $(\phi_i, C_i, T_i, D_i)$, where $\phi_i$, $C_i$, $T_i$, and $D_i$ represent the offset, the WCET, the period, and the relative deadline respectively.
Relative deadlines are called implicit deadlines when $T_i = D_i$, constrained deadlines when $T_i < D_i$, and arbitrary deadlines when $D_i$ is independent of $T_i$.
The utilization of $v^{td}_i$ is denoted as $u_i$, such that $u_i = C_i / T_i$.
The total utilization $U$ of a multi-rate DAG is defined by $U = \sum_{v^{td}_i \in V}u_i$.

\vspace{-1mm}
\subsection{Chain-based DAG}
\label{ssec: chain_based_dag}
\vspace{-1mm}

\begin{figure}[tb]
    \centering
    \scalebox{0.8}{
        \includegraphics[width=\linewidth]{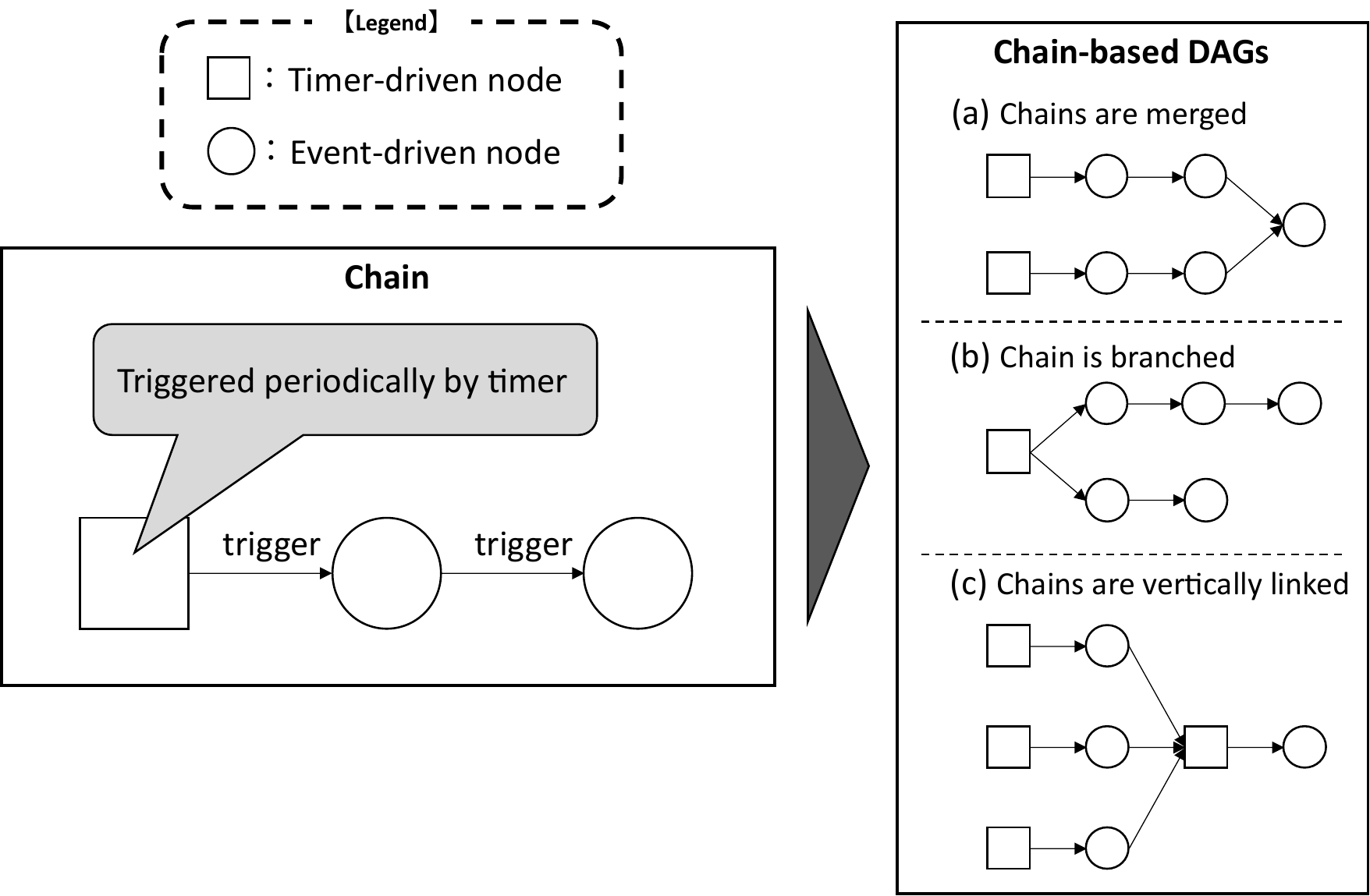}
    }
    \vspace{-3mm}
    \caption{Chain-based DAGs.}
    \label{fig: chain_dag}
    \vspace{-3mm}
\end{figure}

A chain-based DAG is a DAG consisting of chains of multiple event-driven nodes connected to a head timer-driven node, as shown in the left side of Fig.~\ref{fig: chain_dag}.
There are two types of nodes in the chain-based DAG: (i) timer-driven nodes that are triggered at predetermined periods, denoted by $v^{td}_i$, and (ii) event-driven nodes that are triggered by receiving data from predecessor nodes, denoted by $v^{ed}_i$.
The definition of timer-driven nodes is the same as in Section~\ref{ssec: multi_rate_dag}.

Each chain $\Gamma_i$ is denoted as $\Gamma_i = \{v^{td}_i, v^{ed}_k, ..., v^{ed}_{|\Gamma_i|}\}$, where $|\Gamma_i|$ is the number of nodes that compose $\Gamma_i$.
The head $v^{td}_i$ in the chain is triggered periodically, and subsequent event-driven nodes $v^{ed}_k$ are triggered by their direct predecessors.
This definition is the same as that used in existing studies~\cite{tang2020response}.
The WCET of $\Gamma_i$ is denoted by $C_{\Gamma_i}$ and defined by $C_{\Gamma_i} = \sum_{v_j \in \Gamma_i}C_j$.

% \vspace{-4mm}
% \begin{equation}
%     \label{eq: wcet_chain}
%     C_{\Gamma_i} = \sum_{v_j \in \Gamma_i}C_j
% \end{equation}
% \vspace{-3mm}

Because the chain is executed in a manner dependent on the period of the head timer-driven node, the utilization of the chain $u_{\Gamma_i}$ is calculated in $u_{\Gamma_i} = \frac{C_{\Gamma_i}}{T_{\Gamma_i}}$.

% \vspace{-2mm}
% \begin{equation}
%     \label{eq: chain_utilization}
%     u_{\Gamma_i} = \frac{C_{\Gamma_i}}{T_{\Gamma_i}}
% \end{equation}
% \vspace{-3mm}

Here, $T_{\Gamma_i}$ is the period of the head timer-driven node $v^{td}_i$ of the chain.
The total utilization of the chain-based DAG is defined by $U = \sum_{\Gamma_i \in V}u_{\Gamma_i}$.

% \vspace{-4mm}
% \begin{equation}
%     \label{eq: chain_total_utilization}
%     U = \sum_{\Gamma_i \in V}u_{\Gamma_i}
% \end{equation}
% \vspace{-2mm}

Chain-based DAGs primarily exist in robot operating system (ROS)-based and AUTOSAR-adaptive systems~\cite{choi2021picas, yano2021access}.
In a typical ROS-based system, such as a self-driving system, different sensor data are processed and merged by multiple chains to output the final command.
When modeling ROS-based systems as DAGs, it is necessary to consider DAGs in which multiple chains merge ((a) in Fig.~\ref{fig: chain_dag}), where the chain branches ((b) in Fig.~\ref{fig: chain_dag}), and those in which multiple chains are vertically linked ((c) in Fig.~\ref{fig: chain_dag}).
When an event-driven node has multiple predecessor nodes, as in (a) in Fig.~\ref{fig: chain_dag}, the node is triggered when the data from all the predecessor nodes are available.

\vspace{-1mm}
\section{Design and implementation}
\label{sec: design_implementation}
\vspace{-1mm}

This section describes the design and implementation of RD-Gen.
RD-Gen iterates over the construction of the graph and the set of properties for all parameter combinations described by the user in the configuration file as shown in the upper right of Fig.~\ref{fig: system_model}.
RD-Gen is an open-source tool and is publicly available\footnote{https://github.com/azu-lab/RD-Gen}.
% Section~\ref{ssec: common} shows the common functionalities of RD-Gen, Section~\ref{ssec: graph_construction} describes the three graph construction methods of RD-Gen, and Section~\ref{ssec: set_properties} explains how to set the properties for a graph.

\begin{figure}[tb]
    \begin{tabular}{c}
        \begin{minipage}{0.495\linewidth}
            \centering
            \begin{tabular}{l}
                \lstset{linewidth=4.0cm, basicstyle=\scriptsize}
\begin{lstlisting}
Seed: 0
Number of DAGs: 100

Graph structure:
  Generation method: Fan-in/Fan-out
  (*\textbf{Number of nodes:} *)
    (*\textbf{Combination: [10, 20]} *)
  In-degree:
    Random: (start=1, stop=3, step=1)
  Out-degree:
    Random: (1, 3, 1)
  (*\textbf{Number of entry nodes:} *)
    (*\textbf{Combination: [1, 3]} *)
  Number of exit nodes:
    Fixed: 1

Properties:
  Execution time:
    Random: (1, 50, 1)
\end{lstlisting}
            \end{tabular}
        \end{minipage}
        \hfill
        \begin{minipage}{0.495\linewidth}
            \centering
            \scalebox{0.85}{
                \includegraphics[width=\linewidth]{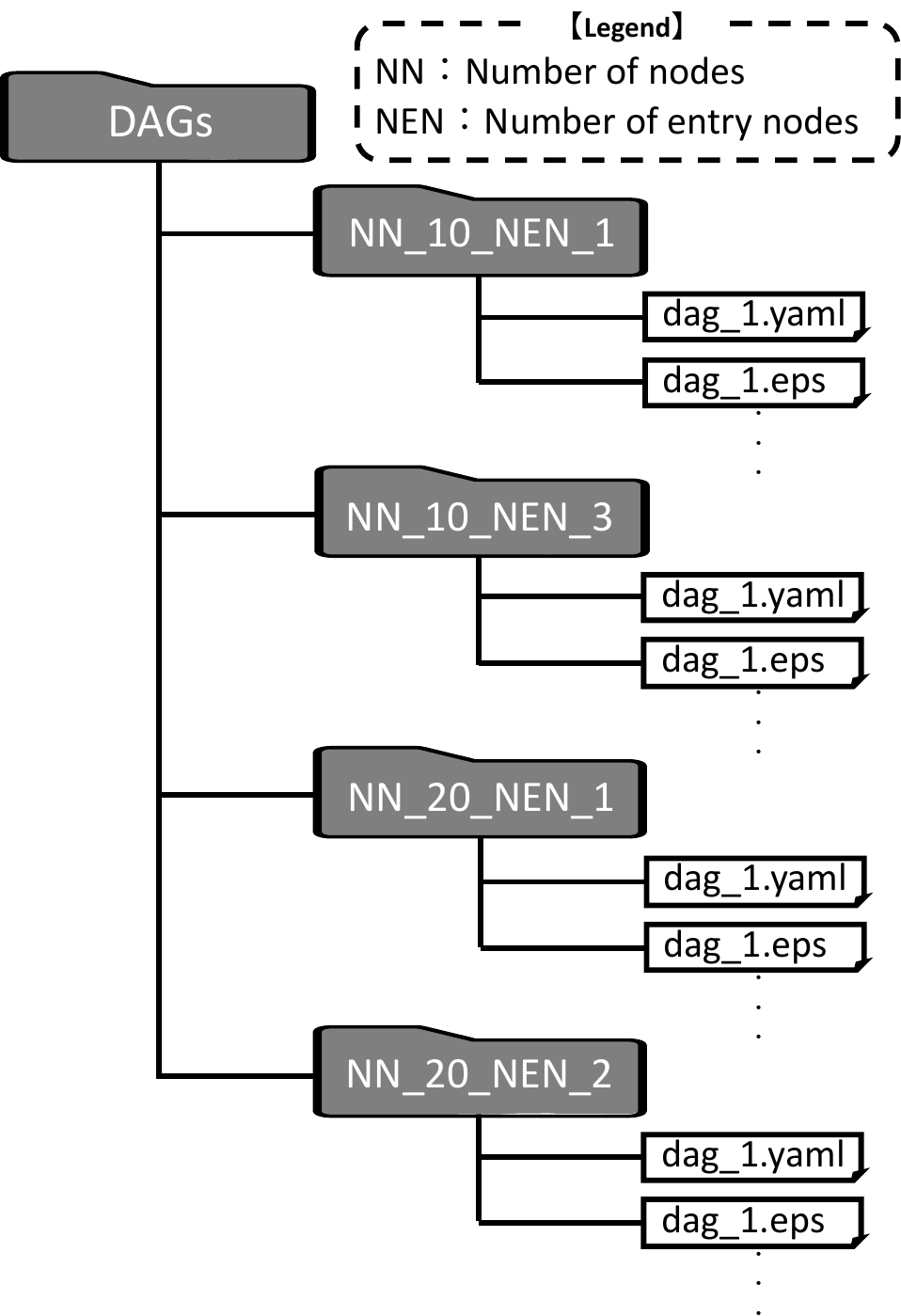}
            }
        \end{minipage}
    \end{tabular}
    \vspace{-3mm}
    \caption{Example of generating all combinations.}
    \label{fig: combo_exam}
    \vspace{-7mm}
\end{figure}

\begin{table*}[tb]
    \vspace{-3mm}
    \caption{All parameters of the {\it Graph structure} in RD-Gen}
    \label{tab: graph_structure}
    \vspace{-3mm}
    \renewcommand{\arraystretch}{1.0}
    \centering
    \scalebox{0.85}{
        \begin{tabular}{|l|lll|}
            \hline
            \multicolumn{1}{|c|}{\textbf{Generation methods}}                                                                            & \multicolumn{2}{c|}{\textbf{Parameters}}         & \multicolumn{1}{c|}{\textbf{Descriptions}}                                                                                                                  \\ \hline
            \MC{1}{|l|}{\multirow{4}{*}{\begin{tabular}[c]{@{}l@{}}{\it \textbf{Fan-in/Fan-out}}\\ {\it \textbf{G(n, p)}}\end{tabular}}} & \MC{2}{l|}{{\it Number of nodes}}                & Number of nodes in a single DAG                                                                                                                             \\ \cline{2-4}
                                                                                                                                         & \MC{2}{l|}{{\it Number of source nodes}}         & Number of source nodes in a single DAG                                                                                                                      \\ \cline{2-4}
            \MC{1}{|l|}{}                                                                                                                & \MC{2}{l|}{{\it Number of sink nodes}}           & Number of sink nodes in a single DAG                                                                                                                        \\ \cline{2-4}
            \MC{1}{|l|}{}                                                                                                                & \MC{2}{l|}{{\it Ensure weakly connected}}        & When True is specified, the generated DAGs are always weakly connected                                                                                      \\ \hline
            \MC{1}{|l|}{\MR{2}{{\it \textbf{Fan-in/Fan-out}}}}                                                                           & \MC{2}{l|}{{\it In-degree}}                      & Number of edges input to one node                                                                                                                           \\ \cline{2-4}
            \MC{1}{|l|}{}                                                                                                                & \MC{2}{l|}{{\it Out-degree}}                     & Number of edges output from one node                                                                                                                        \\ \hline
            \MC{1}{|l|}{{\it \textbf{G(n, p)}}}                                                                                          & \MC{2}{l|}{{\it Probability of edge existence}}  & Probability of an edge presents between any nodes                                                                                                           \\ \hline
            \MC{1}{|l|}{\MR{12}{{\it \textbf{Chain-based}}}}                                                                             & \MC{2}{l|}{{\it Number of chains}}               & Number of chains in a single chain-based DAG                                                                                                                \\ \cline{2-4}
            \MC{1}{|l|}{}                                                                                                                & \MC{2}{l|}{{\it Main sequence length}}           & Length of the main sequence in a single chain                                                                                                               \\ \cline{2-4}
            \MC{1}{|l|}{}                                                                                                                & \MC{2}{l|}{{\it Number of sub sequences}}        & Number of branches from the main sequence in a single chain                                                                                                 \\ \cline{2-4}
            \MC{1}{|l|}{}                                                                                                                & \MC{1}{l|}{\MR{5}{{\it Vertically link chains}}} & \MC{1}{l|}{{\it Number of source nodes}}                               & Number of source nodes in a single DAG                                             \\ \cline{3-4}
            \MC{1}{|l|}{}                                                                                                                & \MC{1}{l|}{}                                     & \MC{1}{l|}{{\it Main sequence tail}}                                   & \tabml{When True is specified, the tail of the main sequence is randomly connected \\ to the head of another chain} \\ \cline{3-4}
            \MC{1}{|l|}{}                                                                                                                & \MC{1}{l|}{}                                     & \MC{1}{l|}{{\it Sub sequence tail}}                                    & \tabml{When True is specified, the tail of the sub sequence is randomly connected  \\ to the head of another chain}  \\ \cline{2-4}
            \MC{1}{|l|}{}                                                                                                                & \MC{1}{l|}{\MR{4}{{\it Merge chains}}}           & \MC{1}{l|}{{\it Number of sink nodes}}                                 & Number of sink nodes in a single DAG                                               \\ \cline{3-4}
            \MC{1}{|l|}{}                                                                                                                & \MC{1}{l|}{}                                     & \MC{1}{l|}{{\it Middle of chain}}                                      & \tabml{When True is specified, merge from a tail node of a random chain to a node  \\ other than the head and tail of another chain} \\ \cline{3-4}
            \MC{1}{|l|}{}                                                                                                                & \MC{1}{l|}{}                                     & \MC{1}{l|}{{\it Sink node}}                                            & When True is specified, merge a head node of a random chain with the sink node     \\ \hline
        \end{tabular}
    }
    \vspace{-5.5mm}
\end{table*}

\vspace{-1mm}
\subsection{Common Functionality}
\label{ssec: common}
\vspace{-1mm}

RD-Gen receives as input a YAML file written by the user, and with a single command, RD-Gen generates an entire set of random DAGs with different parameter combinations.
An example of an input parameter file in the YAML format is shown on the left side of Fig.~\ref{fig: combo_exam}.
Because RD-Gen allows the user to specify seed values for pseudo-random number generation (the {\it Seed} parameter in Fig.~\ref{fig: combo_exam}), the same random DAG set can be generated by entering the same YAML file into RD-Gen.
{\it Number of DAGs} in Fig.~\ref{fig: combo_exam} indicates the number of DAGs randomly generated for each combination.
The shape of the generated DAGs is determined by the parameters under {\it Graph structure}, and the properties of the nodes and edges are determined by the parameters under {\it Properties}.

For parameters that require a numeric input, users can specify values in the following three ways.
\begin{enumerate}
    \item {\it Fixed}: Only one value is specified, and this value is always used when a DAG is generated. This specification method can be used for DAGs in which end-to-end deadlines are considered, and one sink node is always desired, as in the {\it Number of exit nodes} parameter in Fig.~\ref{fig: combo_exam}.
    \item {\it Random}: When a DAG is generated, one value is randomly selected using uniform distribution from an input range. This is the most basic specification method that is used when there is no special intended value and when it is desirable to have a variety of values, as shown in the {\it In-degree}, {\it Out-degree}, and {\it Execution time} parameters in Fig.~\ref{fig: combo_exam}.
    \item \csix{{\it Combination}: RD-Gen generates DAGs for all combinations of all lists of parameters for which {\it Combination} is specified. This specification method allows all DAG sets used in random evaluations in DAG studies to be generated in single command execution. Therefore, RD-Gen can significantly reduce the time and effort required by researchers to generate DAGs. For example, when {\it Combination} of {\it Number of nodes} is {\it [10, 20]} and {\it Combination} of {\it Number of source nodes} is {\it [1, 3],} RD-Gen generates 100 DAGs for \rr{1.2}{each} combination, as shown on the right side of Fig.~\ref{fig: combo_exam}.}
\end{enumerate}

When specifying ranges for {\it Random} and {\it Combination} in RD-Gen, the following two intuitive descriptions are possible.
\begin{enumerate}
    \item {\it List format}: The user specifies their choices in a list (array) format (such as {\it [10, 20]} for {\it Number of nodes} in Fig.~\ref{fig: combo_exam}).
    \item {\it Tuple format}: The user specifies a {\it start} value, a {\it stop} value, and a {\it step} (such as {\it (start=1, stop=3, step=1)} for {\it In-degree} in Fig.~\ref{fig: combo_exam}). The tuple format is internally expanded into a list format of values that are sequentially added from the {\it start} value to the {\it stop} value at an interval of size {\it step} (e.g., {\it (start=1, stop=3, step=1)} expands to a list of {\it [1, 2, 3]}). Here, {\it \dq{start=,}} {\it \dq{stop=,}} and {\it \dq{step=}} are optional, as in {\it Out-degree} in Fig.~\ref{fig: combo_exam}.
\end{enumerate}

RD-Gen can output DAG description files in the YAML, JSON, XML, and DOT formats.
The DOT format is a plain text format used in Graphviz that allows a simple description of graph structure and properties associated with nodes and edges.
EPS, PDF, SVG, and PNG formats can be specified for the DAG images output by RD-Gen.

\vspace{-1mm}
\subsection{Graph Construction}
\label{ssec: graph_construction}
\vspace{-1mm}

RD-Gen extends the {\it Fan-in/Fan-out}~\cite{tgff} and {\it G(n, p)}~\cite{cordeiro2010random} methods widely used in the scheduling field by DAG researchers.
In addition, RD-Gen provides a {\it Chain-based} method for generating state-of-the-art chain-based DAGs as shown in Fig.~\ref{fig: chain_dag}.
The {\it Chain-based} method is a new method proposed in this paper to generate chain-based DAGs with flexibility.
The parameters that can be specified in the {\it Graph structure} of RD-Gen are listed in Table~\ref{tab: graph_structure}.

\begin{table}[tb]
    \centering
    \caption{Helper functions in the algorithms}
    \label{tab: helper_functions}
    \vspace{-3mm}
    \renewcommand{\arraystretch}{1.2}
    \scalebox{0.85}{
        \begin{tabular}{|c|l|}\hline
            \textbf{Symbols}                           & \MC{1}{c|}{\textbf{Descriptions}}                       \\\hline
            Random($a$, $b$) $|$ $a, b \in \mathbb{N}$ & \tabml{Function that returns a random natural           \\ number between $a$ and $b$.} \\ \hline
            In($v_i$)                                  & Function that returns the in-degree of $v_i$            \\\hline
            Out($v_i$)                                 & Function that returns the out-degree of $v_i$           \\\hline
            Sequence($\{v_i, ..., v_k\}$)              & \tabml{Function that returns a linear sequence of nodes \\ connected by edges consisting of input nodes. \\ For example, when the input is $\{v_1, v_2, v_3\}$, \\ it returns $\{v_1, e_{1, 2}, v_2, e_{2, 3}, v_3\}$.} \\\hline
        \end{tabular}
    }
    \vspace{1mm}
\end{table}

\setlength{\textfloatsep}{-3pt}% Remove \textfloatsep
\setlength{\floatsep}{0pt}
\begin{algorithm}[t]
    \linespread{0.9}\selectfont
    {\footnotesize
        \KwIn{
            $n$ \la {\it Number of nodes} \\
            $nsnk$ \la {\it Number of source nodes} \\
            $nsrc$ \la {\it Number of sink nodes} \\
            $id$ \la {\it In-degree} \\
            $od$ \la {\it Out-degree} \\
        }
        \KwOut{A DAG that satisfies user-specified parameters}
        Initialize $G$ \la $(V, E)$, with $V$ \la $\{v_1, ..., v_{nsnk}\}$ and $E$ \la $\{\emptyset\}$ \\
        \While{$|V| \neq n - nsrc$}{

            \If{Random(0, 1) = 1}{
                \tcc{Fan-in phase}
                $S$ \la Set of nodes in $V$ whose out-degree is less than or equal to $od$ \\
                $r$ \la Random(1, $id$) \\
                $T$ \la Randomly choose $r$ nodes from $S$ \\
                $V$ \la $V \cup \{v_{|V|+1}\}$ \\
                \ForEach{$v_i$ $\in$ $T$}{
                    $E$ \la $E \cup \{e_{i, |V|}\}$ \\
                }
            }
            \Else{
                \tcc{Fan-out phase}
                $v_i$ \la The node with the largest difference between $od$ and its out-degree in $V$ \\
                $diff$ \la $od$ $-$ Out($v_i$) \\
                $r$ \la Random(1, $diff$) \\
                \For{$k$ \la $|V|+1$ \KwTo $|V|+r+1$}{
                    $V$ \la $V \cup \{v_{k}\}$ \\
                    $E$ \la $E \cup \{e_{i, k}\}$ \\
                }
            }
            \If{$|V| > n - nsrc$}{
                Initialize $G$ \la $(V, E)$, with $V$ \la $\{v_1, ..., v_{nsnk}\}$ and $E$ \la $\{\emptyset\}$
            }
        }
        $G$ \la add\_sink\_nodes($G$, $nsrc$) \\
        $G$ \la weakly\_connect($G$) \\
        return $G$
        \caption{{\it Fan-in/Fan-out} method in RD-Gen}
        \label{alg: fan_in_fan_out}
    }
\end{algorithm}

\begin{algorithm}[t]
    \linespread{0.9}\selectfont
    {\footnotesize
    \KwIn{$G$: A DAG, $nsrc$: {\it Number of sink nodes}}
    \KwOut{DAG with sink nodes added}
    $NSRC$ \la $\{v_{|V|+1}, ..., v_{|V|+nsrc+1}\}$ \\
    $CEX$ \la Set of current sink nodes of $G$ \\
    \While{$\exists n \in NSRC$, In($v_{n}) = 0$ $\vee$ $\exists c \in CEX$, out($v_{c}) = 0$}{
        $v_i$ \la The node with the smallest out-degree in $CEX$ \\
        $v_j$ \la The node with the smallest in-degree in $NSRC$ \\
        $E$ \la $E \cup \{e_{i, j}\}$ \\
    }
    $V$ \la $V \cup NSRC$ \\
    return $G$
    \caption{add\_sink\_nodes($G$, $nsrc$)}
    \label{alg: add_exit_nodes}
    }
\end{algorithm}

\begin{algorithm}[t]
    \linespread{0.9}\selectfont
    {\footnotesize
        \KwIn{$G$: A DAG}
        \KwOut{A weakly connected DAG}
        \While{$G$ is not weakly connected}{
            $MC$ \la Weakly connected component with the maximum number of nodes in $G$ \\
            $RC$ \la Randomly selected a weakly connected component other than $MC$ \\
            $v_i$ \la Randomly choose one of the sink nodes in $RC$ \\
            $v_j$ \la Randomly choose one node other than source nodes in $MC$ \\
            $E$ \la $E \cup \{e_{i, j}\}$ \\
        }
        return $G$
        \caption{weakly\_connect($G$)}
        \label{alg: weakly_connect}
    }
\end{algorithm}

\subsubsection{Fan-in/Fan-out Method}
\label{sssec: fan_in_fan_out}

{\it Fan-in/Fan-out} is a random DAG generation method proposed by Dick et al. that is provided by TGFF~\cite{tgff}.
The {\it Fan-in/Fan-out} method can specify the in-degree (i.e., the number of edges to be input) and out-degree (i.e., the number of edges to be output) ranges for a single node; a DAG is then generated in which all nodes satisfy this condition.
The {\it Fan-in/Fan-out} method extends the graph by randomly repeating the Fan-in and Fan-out phases.
However, the original {\it Fan-in/Fan-out} method cannot completely satisfy the requirements of researchers using DAGs.

The DAG generated by the {\it Fan-in/Fan-out} method has just one source node, while in-vehicle and self-driving systems have multiple sensors~\cite{verucchi2020latency} and require a DAG with multiple source nodes.
Even though the latest version of TGFF allows multiple source nodes, it may generate DAGs that are not weakly connected.
Weakly connected graphs are graphs that \rr{3.6}{any two nodes are reachable} without regard to the orientation of the edges, and DAGs that are not weakly connected are separated.
Therefore, such DAGs may cause unintended evaluation results and must be avoided.
It is also necessary to be able to specify the number of sink nodes because many studies consider DAGs with a single sink node~\cite{cho2021conditionally, zhang2020efficient}.
In addition, because scheduling and allocation methods using DAGs affect the performance depending on the number of nodes in a DAG, evaluations are performed with various changes in the number of nodes~\cite{senapati2021hmds}.
However, TGFF does not allow a user to completely control the number of nodes to be generated in a DAG.

To meet these requirements, RD-Gen allows the specification of the number of nodes, source nodes, and sink nodes in a single DAG and ensures that the DAG is weakly connected.
The {\it Fan-in/Fan-out} method procedure provided by RD-Gen is shown in Algorithm~\ref{alg: fan_in_fan_out}, and the helper functions used in the algorithms are listed in Table~\ref{tab: helper_functions}.
First, the DAG is initialized with a specified number of source nodes (line 1 in Algorithm~\ref{alg: fan_in_fan_out}).
RD-Gen then iteratively expands the graph until the number of nodes fits within a user-specified value (line 2 in Algorithm~\ref{alg: fan_in_fan_out}).
Next, the graph is randomly extended in the Fan-in phase or Fan-out phase (lines 3--20 in Algorithm~\ref{alg: fan_in_fan_out}).
If the number of nodes exceeds the user-specified value, the DAG is initialized (lines 21--23 in Algorithm~\ref{alg: fan_in_fan_out}).
After the loop ends, the {\it add\_sink\_nodes} function is called, and nodes with no output edges are merged into a user-specified number of sink nodes with the minimum number of edges (Algorithm~\ref{alg: add_exit_nodes}).
Finally, the {\it weakly\_connect} function is called to add edges until the DAG is weakly connected (Algorithm~\ref{alg: weakly_connect}).
With these extensions, the {\it Fan-in/Fan-out} method implemented by RD-Gen allows the user to fully control the number of nodes, source nodes, and sink nodes in a single DAG and guarantees that the DAG is weakly connected.

\begin{algorithm}[t]
    \linespread{0.9}\selectfont
    {\footnotesize
        \KwIn{
            $n$ \la {\it Number of nodes} \\
            $p$ \la {\it probability of edge existence} \\
            $nsnk$ \la {\it Number of source nodes} \\
            $nsrc$ \la {\it Number of sink nodes} \\
        }
        \KwOut{A DAG that satisfies user-specified parameters}
        $n$ \la $n - nsnk - nsrc$ \\
        Initialize $G$ \la $(V, E)$, with $V$ \la $\{v_1, ..., v_{n}\}$ and $E$ \la $\{\emptyset\}$ \\
        \For{$i$ \la $1$ \KwTo $|V|$}{
            \For{$j$ \la $1$ \KwTo $|V|$}{
                \If{(Random(0, 100) $\le p*100$) $\wedge$ $i < j$}{
                    $E$ \la $E \cup \{e_{i, j}\}$
                }
            }
        }
        \tcc{Add source nodes}
        $NSNK$ \la $\{v_{|V|+1}, ..., v_{|V|+nsnk+1}\}$ \\
        $CEN$ \la Set of current source nodes of $G$ \\
        \While{$\exists n \in NSNK$, Out($v_n) = 0$ $\vee$ $\exists c \in CEN$, in($v_c) = 0$}{
            $v_i$ \la The node with the smallest in-degree in $CEN$ \\
            $v_j$ \la The node with the smallest out-degree in $NSNK$ \\
            $E$ \la $E \cup \{e_{i, j}\}$ \\
        }
        $V$ \la $V \cup NSNK$ \\
        $G$ \la add\_sink\_nodes($G$, $nsrc$) \\
        $G$ \la weakly\_connect($G$) \\
        return $G$
        \caption{{\it G(n, p)} method in RD-Gen}
        \label{alg: g_n_p}
    }
\end{algorithm}

\subsubsection{G(n, p) Method}
\label{sssec: g_n_p}

{\it G(n, p)} is a well-known graph construction method proposed by Paul Erd{\H{o}}s, and Alfr{\'e}d R{\'e}nyi et al.~\cite{cordeiro2010random} that constructs a graph according to the number of nodes and the probability of an edge existing between any two nodes.
Many of the random DAG sets used in evaluations in the latest studies using DAGs have been constructed based on the {\it G(n, p)} method~\cite{voronov2021ai, he2021response, dong2019efficient}.
However, there are problems with the original {\it G(n, p)} method, such as the possibility of a cycle in a graph and the inability to guarantee that the graph is weakly connected.

Accordingly, in RD-Gen, the {\it G(n, p)} method is extended for direct use in the evaluation of DAG studies.
The procedure for the {\it G(n, p)} method in RD-Gen is shown in Algorithm~\ref{alg: g_n_p}.
First, nodes other than the source and sink nodes are added to the graph (lines 1 and 2 in Algorithm~\ref{alg: g_n_p}).
Then, for any two nodes, the addition of an edge is tried with \rr{1.3}{50\% probability}.
Here, if the index of the destination node is smaller than the index of the source node, the edge addition is canceled (line 5 in Algorithm~\ref{alg: g_n_p}).
This condition results in no cycles being present in the graph~\cite{voronov2021ai}.
After the loop ends, a user-specified number of source nodes and a node with an in-degree of zero are connected with the smallest edge.
Finally, the {\it add\_sink\_nodes} and {\it weakly\_connect} functions are called in the same manner as in the {\it Fan-in/Fan-out} method.

\begin{algorithm}[t]
    \linespread{0.9}\selectfont
    {\footnotesize
        \KwIn{
            $nc$ \la {\it Number of chains} \\
            $ml$ \la {\it Main sequence length} \\
            $ns$ \la {\it Number of sub sequences} \\
            $nsnk$ \la {\it Number of source nodes} \\
            $nsrc$ \la {\it Number of sink nodes} \\
        }
        \KwOut{A DAG that satisfies user-specified parameters}
        Initialize $G$ \la $(V, E)$, with $V$ \la $\{\emptyset\}$ and $E$ \la $\{\emptyset\}$ \\
        \tcc{Construct each chain}
        \For{$i$ \la $1$ \KwTo $nc$}{
            $main$ \la Sequence($\{v_{|V|+1}, ..., v_{|V|+ml+1}\}$) \\
            $G$ \la $main$ \\
            \For{$j$ \la $1$ \KwTo $ns$}{
                $r$ \la Random($|V|+1, |V|+ml$) \\
                $sl$ \la Random($1, ml-r$) \\
                $sub$ \la Sequence($\{v_{|V|+1}, ..., v_{|V|+sl+1}\}$) \\
                $E$ \la $E \cup \{e_{r, |V|+1}\}$ \\
                $G$ \la $sub$ \\
            }
        }
        \tcc{Vertically link chains}
        \While{Number of source nodes of $G$ $\neq$ $nsnk$}{
            $v_i$ \la Randomly choose one of the sink nodes in $V$ \\
            $v_j$ \la Randomly choose one node from the head of the chains \\
            $E$ \la $E \cup \{e_{i, j}\}$ \\
        }
        \tcc{Merge chains}
        \While{Number of sink nodes of $G$ $\neq$ $nsrc$}{
            $v_i$ \la Randomly choose one of the sink nodes in $V$ \\
            $v_j$ \la Randomly choose one node from other than the head of the chains \\
            $E$ \la $E \cup \{e_{i, j}\}$ \\
        }
        return $G$
        \caption{{\it Chain-based} method in RD-Gen}
        \label{alg: chain_based}
    }
\end{algorithm}

\begin{figure*}[h]
    \hspace{-3mm}
    \begin{tabular}{l}
        \begin{minipage}[t]{0.7\linewidth}
            \vspace{-40mm}
            \tblcaption{All parameters of {\it Properties} in RD-Gen}
            \label{tab: properties}
            \vspace{-3mm}
            \renewcommand{\arraystretch}{1.0}
            \centering
            \scalebox{0.7}{
                \begin{tabular}{|llll|l|}
                    \hline
                    \multicolumn{4}{|c|}{\textbf{Parameters}}                 & \multicolumn{1}{c|}{\textbf{Descriptions}}                                                                                                                                                  \\ \hline
                    \MC{4}{|l|}{{\it \textbf{Execution time}}}                & Execution time of nodes                                                                                                                                                                     \\ \hline
                    \MC{4}{|l|}{{\it \textbf{Communication time}}}            & Communication time of edges                                                                                                                                                                 \\ \hline
                    \MC{4}{|l|}{{\it \textbf{CCR}}}                           & CCR of a single DAG                                                                                                                                                                         \\ \hline
                    \MC{2}{|l|}{{\it \textbf{End-to-end deadline}}}           & \MC{2}{l|}{{\it Ratio of deadline to critical path}} & Ratio of the end-to-end deadline to the critical path of DAG                                                                         \\ \hline
                    \MC{2}{|l|}{\MR{7}{{\it \textbf{Multi-rate}}}}            & \MC{2}{l|}{{\it Periodic type}}                      & Specify a group of nodes to be timer driven                                                                                          \\ \cline{3-5}
                    \MC{2}{|l|}{}                                             & \MC{2}{l|}{{\it Period}}                             & Period of timer-driven nodes                                                                                                         \\ \cline{3-5}
                    \MC{2}{|l|}{}                                             & \MC{2}{l|}{{\it Source node period}}                 & Period of source timer-driven nodes                                                                                                  \\ \cline{3-5}
                    \MC{2}{|l|}{}                                             & \MC{2}{l|}{{\it Sink node period}}                   & Period of exit timer-driven nodes                                                                                                    \\ \cline{3-5}
                    \MC{2}{|l|}{}                                             & \MC{2}{l|}{{\it Offset}}                             & Offset of timer-driven nodes                                                                                                         \\ \cline{3-5}
                    \MC{2}{|l|}{}                                             & \MC{2}{l|}{{\it Total utilization}}                  & Total utilization of a single DAG                                                                                                    \\ \cline{3-5}
                    \MC{2}{|l|}{}                                             & \MC{2}{l|}{{\it Maximum utilization}}                & Maximum utilization of one timer-driven node                                                                                         \\ \hline
                    \MC{2}{|l|}{\MR{2}{{\it \textbf{Additional properties}}}} & \MC{2}{l|}{{\it Node properties}}                    & User-defined numeric parameters with any name for nodes                                                                              \\ \cline{3-5}
                    \MC{2}{|l|}{}                                             & \MC{2}{l|}{{\it Edge properties}}                    & User-defined numeric parameters with any name for edges                                                                              \\ \hline
                    \MC{2}{|l|}{\MR{9}{{\it \textbf{Output formats}}}}        & \MC{1}{l|}{\multirow{4}{*}{{\it DAG}}}               & {\it YAML}                                                   & \MR{4}{Outputs DAG description files in the format specified by True} \\ \cline{4-4}
                    \multicolumn{2}{|l|}{}                                    & \multicolumn{1}{l|}{}                                & {\it JSON}                                                   &                                                                       \\ \cline{4-4}
                    \multicolumn{2}{|l|}{}                                    & \multicolumn{1}{l|}{}                                & {\it XML}                                                    &                                                                       \\ \cline{4-4}
                    \multicolumn{2}{|l|}{}                                    & \multicolumn{1}{l|}{}                                & {\it DOT}                                                    &                                                                       \\ \cline{3-5}
                    \MC{2}{|l|}{}                                             & \MC{1}{l|}{\MR{5}{{\it Figure}}}                     & {\it Draw legend}                                            & If True, a legend is drawn on the output DAG figure                   \\ \cline{4-5}
                    \multicolumn{2}{|l|}{}                                    & \multicolumn{1}{l|}{}                                & {\it PNG}                                                    & \MR{4}{Outputs DAG figures in the format specified by True}           \\ \cline{4-4}
                    \multicolumn{2}{|l|}{}                                    & \multicolumn{1}{l|}{}                                & {\it SVG}                                                    &                                                                       \\ \cline{4-4}
                    \multicolumn{2}{|l|}{}                                    & \multicolumn{1}{l|}{}                                & {\it EPS}                                                    &                                                                       \\ \cline{4-4}
                    \multicolumn{2}{|l|}{}                                    & \multicolumn{1}{l|}{}                                & {\it PDF}                                                    &                                                                       \\ \hline
                \end{tabular}
            }
        \end{minipage}
            \hspace{0mm}
        \begin{minipage}[c]{0.3\linewidth}
            \vspace{10mm}
            \scalebox{0.95}{
                \includegraphics[width=\linewidth]{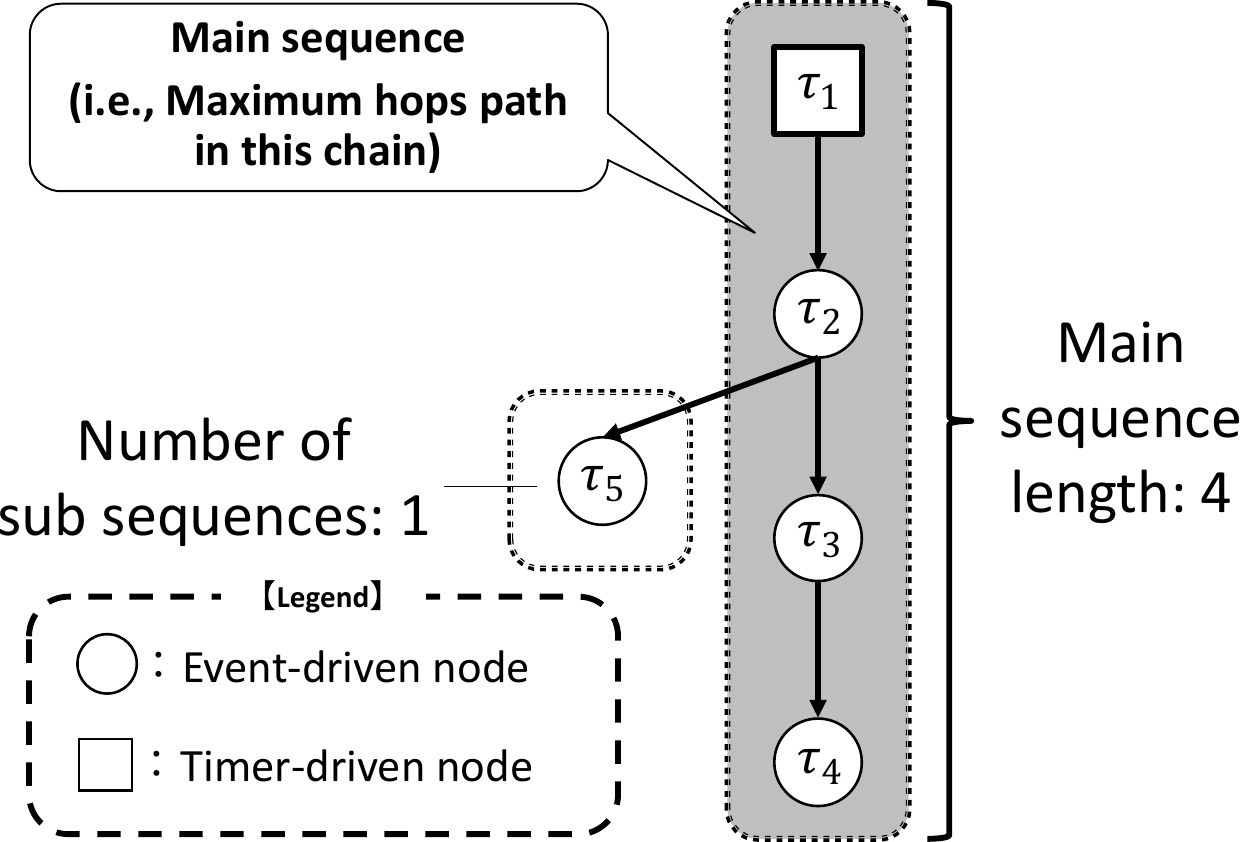}
            }
            \caption{Example of parameters that\\ determine the shape of a single chain.}
            \label{fig: chain_param_exam}
        \end{minipage}
    \end{tabular}
    \vspace{-5mm}
\end{figure*}

\subsubsection{Chain-based Method}
\label{sssec: chain_based_method}

The {\it Chain-based} method designed is a new random DAG generation method proposed in this paper to meet the requirements of modern chain-based DAGs.
The {\it Chain-based} method can easily and reproducibly generate chain-based random DAGs used to evaluate existing studies~\cite{choi2021picas, tang2020response}.
In the {\it Chain-based} method, multiple chains are combined to construct a single DAG.
The parameters that determine the shape of a single chain are explained here using Fig.~\ref{fig: chain_param_exam}.
The user can determine the shape of a given chain by specifying the {\it main sequence length} and the {\it Number of subsequences}.
The main sequence is the path that has the largest number of hops from the head timer-driven node to the tail event-driven node in a single chain ($\{v_1, v_2, v_3, v_4\}$ in Fig.~\ref{fig: chain_param_exam}), and subsequences are straight lines of nodes branching off from the main sequence ($\{v_5\}$ in Fig.~\ref{fig: chain_param_exam}).

In the {\it Chain-based} method, two different ways of connecting chains can be specified.
The first is to link chains vertically, i.e., the event-driven node at the end of each chain is randomly connected to the timer-driven node at the head of another chain (e.g., (c) in Fig.~\ref{fig: chain_dag}).
The second method is to merge multiple chains into one specific node.
The {\it Chain-based} method allows nodes other than the head node of the chain to be specified as integration nodes, and the event-driven node at the tail of each chain is merged into a randomly selected integration node (e.g., (a) in Fig.~\ref{fig: chain_dag}).

The procedure for the {\it Chain-based} method is shown in Algorithm~\ref{alg: chain_based}.
First, chains are constructed for the {\it Number of chains} specified by the user (lines 2--12 in Algorithm~\ref{alg: chain_based}).
In the construction of each chain, the main sequence is generated (lines 3 and 4 in Algorithm~\ref{alg: chain_based}).
Subsequences branch from random nodes other than the tail of the main sequence and are coordinated to not exceed the length of the main sequence (lines 6--10 in Algorithm~\ref{alg: chain_based}).
After all the chains are constructed, the chains are linked vertically until the number of source nodes in the DAG is equal to the user specification (lines 13--17 in Algorithm~\ref{alg: chain_based}).
Finally, the chain is randomly merged until the number of sink nodes in the DAG matches the specified number (lines 18--22 in Algorithm~\ref{alg: chain_based}).
Consequently, the {\it Chain-based} method can flexibly construct DAGs consisting of multiple chains.

\begin{algorithm}[t]
    \linespread{0.9}\selectfont
    {\footnotesize
        \KwIn{
            $G$ \la {DAG} \\
            $pt$ \la {\it Periodic type} \\
            $U$ \la {\it Total utilization} \\
            $p$ \la {\it Period} \\
        }
        \KwOut{A DAG that satisfies user-specified parameters}
        \If{$pt = ``All"$}{
            $u[1, ..., |V|]$ \la UUniFast($|V|, U$)~\cite{bini2005measuring} \\
            \For{$i$ \la $1$ \KwTo $|V|$}{
                $T_i$ \la $p$ \\
                $C_i$ \la $u[i] \times T_i$ \\
            }
        }
        \ElseIf{$pt = ``Chain"$}{
            $u[1, ..., |\Gamma|]$ \la UUniFast($|\Gamma|, U$) \\
            \For{$i$ \la $1$ \KwTo $|\Gamma|$}{
                $T_{\Gamma_i}$ \la $p$ \\
                $C_{\Gamma_i}$ \la $u[i] \times T_{\Gamma_i}$ \\
                $g$ \la Number of nodes in $\Gamma_i$ \\
                $c[i, ..., i+g]$ \la Grouping randomly into $g$ with total equal to $C_{\Gamma_i}$ \\
                \For{$j$ \la $i$ \KwTo $i+g$}{
                    $C_j$ \la $c[j]$
                }
            }
        }
        return $G$
        \caption{Set properties based on the total utilization}
        \label{alg: set_utilization}
    }
\end{algorithm}

\begin{figure*}[t]
    \vspace{-3mm}
    \begin{tabular}{c}
        %1
        \begin{minipage}{0.35\linewidth}
            \begin{flushleft}
                \begin{tabular}{l}
                    \lstset{linewidth=5.5cm, basicstyle=\scriptsize}
\begin{lstlisting}
Seed: 0
Number of DAGs: 100

Graph structure:
  Generation method: Fan-in/Fan-out
  (*\textbf{Number of nodes:} *)
    (*\textbf{Combination: (10, 1000, 10)} *)
  In-degree:
    Random: [1, 2, 3]
  Out-degree:
    Random: [1, 2, 3]
  Number of entry nodes:
    Random: [1, 2, 3, 4, 5]
  Number of exit nodes:
    Fixed: 1
  Ensure weakly connected: True

Properties:
  (*\textbf{Execution time:} *)
    (*\textbf{Random: (1, 30, 1)} *)
  (*\textbf{CCR:} *)
    (*\textbf{Combination: [0.1, 0.2, 0.5, 1.0, 2.0, 5.0, 10.0]} *)
\end{lstlisting}
                \end{tabular}
            \end{flushleft}
            \centering
            Input YAML file
        \end{minipage}
        % 2
        \begin{minipage}{0.22\linewidth}
            \centering
            \scalebox{0.9}{\includegraphics[keepaspectratio, width = \linewidth]{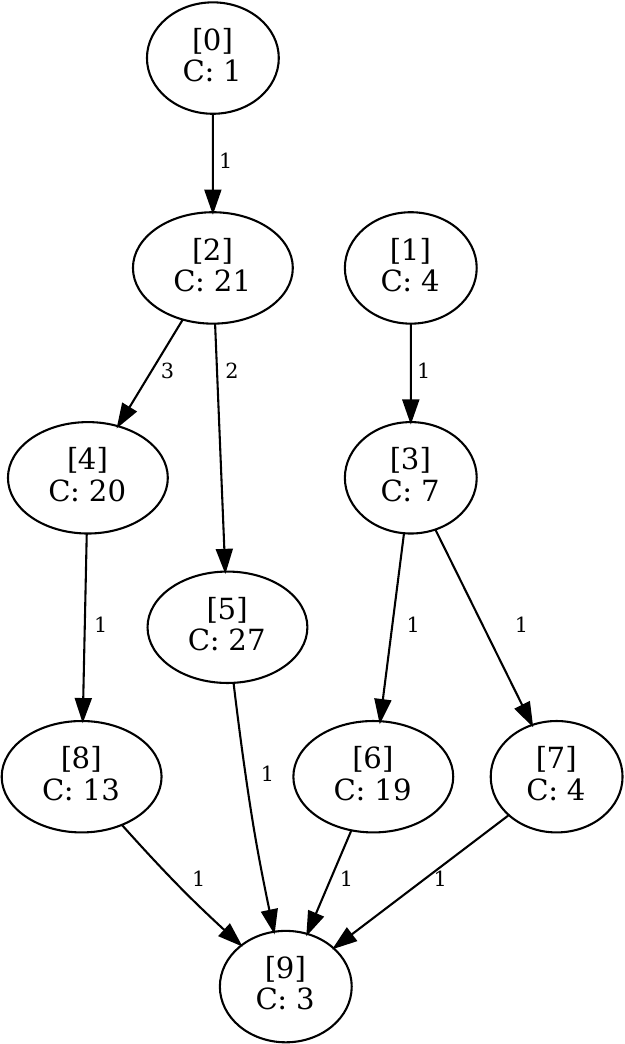}}
            \hspace{3.0cm} Number of nodes: 10, CCR: 0.1
        \end{minipage}
        % 3
        \begin{minipage}{0.47\linewidth}
            \centering
            \scalebox{0.7}{\includegraphics[keepaspectratio, width = \linewidth]{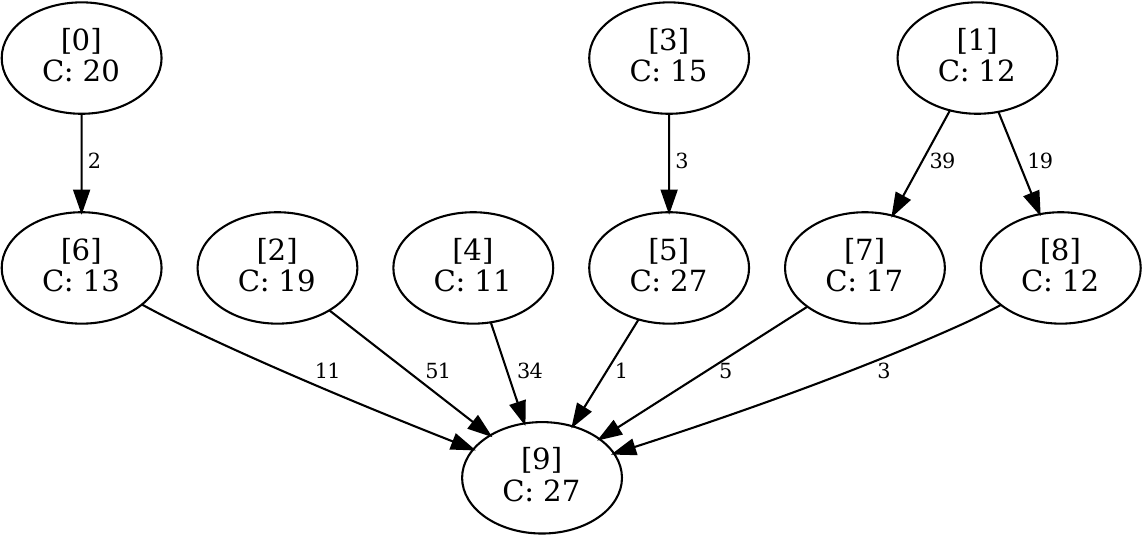}}
            \hspace{3.5cm} Number of nodes: 10, CCR: 1.0 \\
            \vspace{3mm}
            \centering
            \scalebox{0.6}{\includegraphics[keepaspectratio, width = \linewidth]{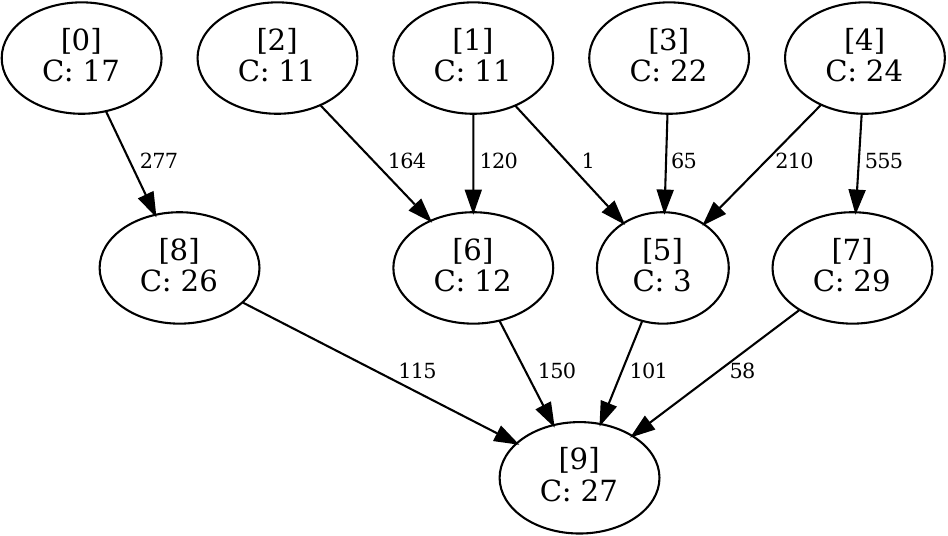}}
            \hspace{3.0cm} Number of nodes: 10, CCR: 10.0
        \end{minipage}
    \end{tabular}
    \centering
    \vspace{-3mm}
    \caption{Example of generating a random DAG set with varying CCR, the number of nodes, the execution time, and the communication time using the {\it Fan-in/Fan-out} method.}
    \label{fig: case_study_single}
    \vspace{-3mm}
\end{figure*}

\begin{figure*}[t]
    \begin{tabular}{c}
        %1
        \begin{minipage}{0.30\linewidth}
            \begin{flushleft}
                \begin{tabular}{l}
                    \lstset{linewidth=4.0cm, basicstyle=\scriptsize}
\begin{lstlisting}
Seed: 0
Number of DAGs: 100

Graph structure:
  Generation method: G(n, p)
  Number of nodes:
    Random: (10, 100, 10)
  Probability of edge:
    Random: (0.1, 0.9, 0.1)
  Number of entry nodes:
    Random: [1, 2, 3, 4, 5]
  Number of exit nodes:
    Random: [1, 2, 3, 4, 5]
  Ensure weakly connected: True

Properties:
  Multi-rate:
    (*\textbf{Periodic type: All} *)
    (*\textbf{Period:} *)
      (*\textbf{Random: (1, 100, 1)} *)
    (*\textbf{Total utilization:} *)
      (*\textbf{Combination: (0.05, 0.95, 0.05)} *)
\end{lstlisting}
                \end{tabular}
            \end{flushleft}
            \centering
            Input YAML file
        \end{minipage}
        % 2
        \begin{minipage}{0.28\linewidth}
            \centering
            \scalebox{0.8}{\includegraphics[keepaspectratio, width = \linewidth]{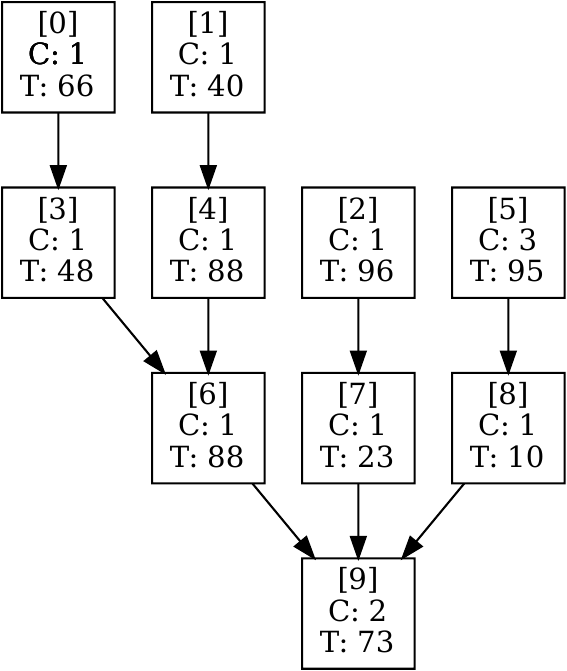}}
            \hspace{3.0cm} Number of nodes: 10,\\ Total utilization: 0.3
        \end{minipage}
        % 3
        \begin{minipage}{0.45\linewidth}
            \centering
            \scalebox{0.53}{\includegraphics[keepaspectratio, width = \linewidth]{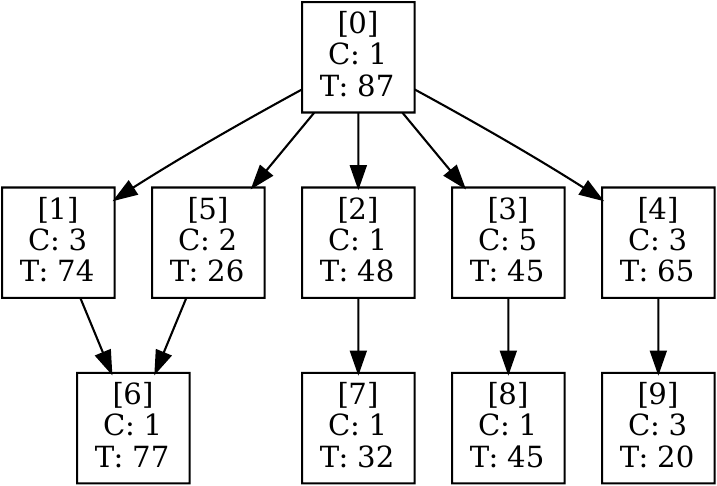}}
            \hspace{3.5cm} Number of nodes: 10, Total utilization: 0.5 \\
            \vspace{3mm}
            \centering
            \scalebox{0.7}{\includegraphics[keepaspectratio, width = \linewidth]{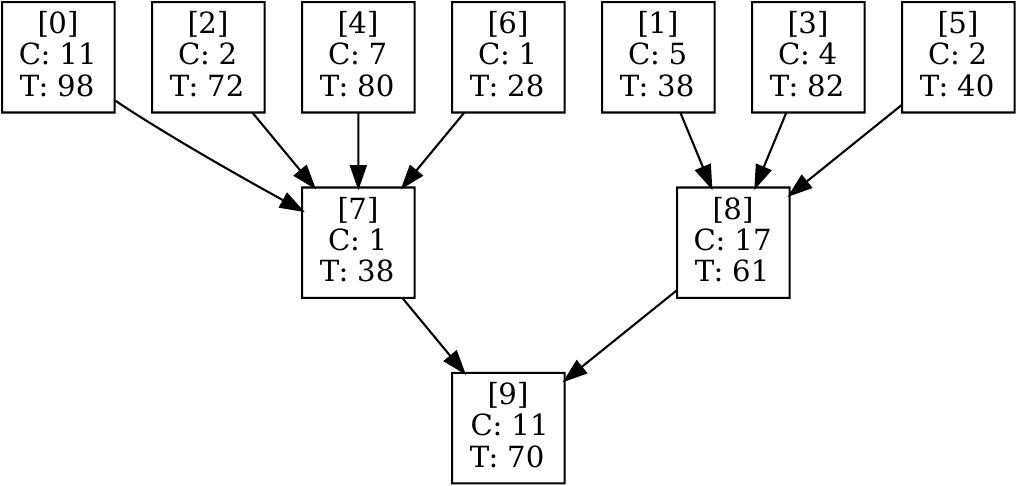}}
            \hspace{3.0cm} Number of nodes: 10, Total utilization: 0.95
        \end{minipage}
    \end{tabular}
    \centering
    \vspace{-3mm}
    \caption{Example of generating multi-rate DAGs of various total utilization.}
    \label{fig: case_study_all_timer}
    \vspace{-6mm}
\end{figure*}

\vspace{-1mm}
\subsection{Property Setting}
\label{ssec: set_properties}
\vspace{-1mm}

RD-Gen automatically sets the typical properties that characterize the nodes and edges of a DAG to meet the user requirements. All of the properties that can be set automatically in RD-Gen are listed in Table~\ref{tab: properties}.
RD-Gen can generate (i) single-rate DAGs, (ii) multi-rate DAGs, and (iii) chain-based DAGs depending on the {\it Periodic type} parameter.
If the {\it Periodic type} is not specified, DAGs of type (i) are generated; if the Periodic type is {\it \dq{All}} DAGs of type (ii) are generated, and if the {\it Periodic type} is {\it \dq{Chain}} DAGs of type (iii) are generated.
    {\it Periodic type} ``Chain'' can only be used in the {\it chain-based} method.
Here, even in a graph constructed by a {\it chain-based} method, all nodes can be timer-driven nodes if {\it Periodic type} ``All'' is specified.
This design is intended to improve the flexibility of RD-GEN.

In RD-Gen, all properties described in Section~\ref{sec: system_model} can be specified using {\it \dq{Fixed,}} {\it \dq{Random,}} or {\it \dq{Combination.}}
Even though existing random DAG generation tools such as TGFF and GGen provide the functionality to randomly assign properties, they do not allow the user to control the values calculated by multiple properties such as CCR and the total utilization.
Because CCR and the total utilization have a significant impact on the performance of the scheduling algorithms and the analysis methods, evaluations are performed by varying these values~\cite{he2021response, senapati2021hmds}.
Therefore, RD-Gen also supports the specification of such complex property values.

An example of the property setting based on the total utilization in RD-Gen is shown in Algorithm~\ref{alg: set_utilization}.
RD-Gen uses the UUniFast method~\cite{bini2005measuring} to set the utilization to each node in a uniform distribution (lines 2 and 9 in Algorithm~\ref{alg: set_utilization}).
If the {\it Periodic type} is {\it \dq{All,}} the utilization for each node is determined based on the {\it Total utilization} specified by the user, and the period and the execution time are set to satisfy this utilization (lines 1--7 in Algorithm~\ref{alg: set_utilization}).
Here, when {\it Maximum utilization} is specified, RD-Gen allocates utilization to each node not to exceed {\it Maximum utilization}.
To support unique implementations of existing studies~\cite{choi2021picas, tang2020response} with RD-Gen, this paper proposes an extension of the UUniFast method to chains.
If the {\it Periodic type} is {\it \dq{Chain,}} the utilization of each chain is determined based on the {\it Total utilization}, and the period of each chain and the sum of the execution time are calculated accordingly (lines 8--12 in Algorithm~\ref{alg: set_utilization}).
The total execution time is randomly divided by the number of nodes in the chain to set the execution time for each node (lines 13--17 in Algorithm~\ref{alg: set_utilization}).
In this way, RD-Gen randomly and automatically sets properties according to complex parameters specified by the user.

RD-Gen provides other parameters that allow for setting the ratio of the end-to-end deadlines to the critical path length and the period of source and sink nodes.
In addition, users can define unique parameters for simple properties that randomly assign numerical values to nodes or edges.

\vspace{-1mm}
\section{Case Study}
\label{sec: case_study}
\vspace{-1mm}

This section illustrates that RD-Gen can generate the random DAG sets used in the evaluation of existing studies based on DAGs.
Case studies of a single-rate DAG, a multi-rate DAG, and a chain-based DAG are shown, respectively.

\subsubsection{Case Study 1: Single-rate DAGs with Various CCR}
\label{ssec: case_study_1}

Since CCR changes the nature of DAGs and affects the performance of scheduling algorithms, existing studies of single-rate DAGs have used random DAG sets with different CCR values in their evaluations.
An example of the generation in RD-Gen of a random DAG set with varying CCR, the number of nodes, the execution time, and the communication time using the {\it Fan-in/Fan-out} method as used in existing studies~\cite{subbaraj2020multi, liu2016minimizing, sheikh2016sixteen} is shown in Fig.~\ref{fig: case_study_single}.
Because the {\it Combination} is specified for the {\it Number of nodes} and {\it CCR} in Fig.~\ref{fig: case_study_single}, RD-Gen generates 100 random DAGs each for all combinations of these parameters (i.e., $\{10, 20, 30, ..., 1000\} \times \{0.1, 0.2, 0.5, 1.0, 2.0, 5.0, 10.0\}$).
For each DAG, after graph construction, the execution time is randomly assigned to each node in the range of 1 to 30 (the {\it Execution time} parameter in Fig.~\ref{fig: case_study_single}).
Then, the total communication time is calculated from the CCR and the sum of the execution time using Eq.~(\ref{eq: ccr}), and the total communication time is randomly distributed to each edge.
Thus, the user can generate all DAG sets used in the evaluation with a single command, without adjusting the number of nodes and CCR.

\begin{figure*}[t]
    \vspace{-3mm}
    \begin{tabular}{c}
        %1
        \begin{minipage}{0.25\linewidth}
            \begin{flushleft}
                \begin{tabular}{l}
                    \lstset{linewidth=4.5cm, basicstyle=\scriptsize}
\begin{lstlisting}
Seed: 0
Number of DAGs: 100

Graph structure:
  Generation method: Chain-based
  (*\textbf{Number of chains:} *)
    (*\textbf{Random: [2, 3, 4, 5, 6, 7, 8, 9, 10]} *)
  Main sequence length:
    Random: (2, 7, 1)
  Merge chains:
    Number of exit nodes:
      Random: [1, 2, 3, 4, 5]
    Middle of chain: False
    Exit node: True

Properties:
  Multi-rate:
    Periodic type: Chain
    Period:
      Random: (50, 1000, 1)
    (*\textbf{Total utilization:} *)
      (*\textbf{Combination: (0.5, 4.0, 0.5)} *)
    (*\textbf{Maximum utilization:} *)
      (*\textbf{Fixed: 1.0} *)
\end{lstlisting}
                \end{tabular}
            \end{flushleft}
            \centering
            Input YAML file
        \end{minipage}
        % 2
        \begin{minipage}{0.32\linewidth}
            \centering
            \scalebox{0.7}{\includegraphics[keepaspectratio, width = \linewidth]{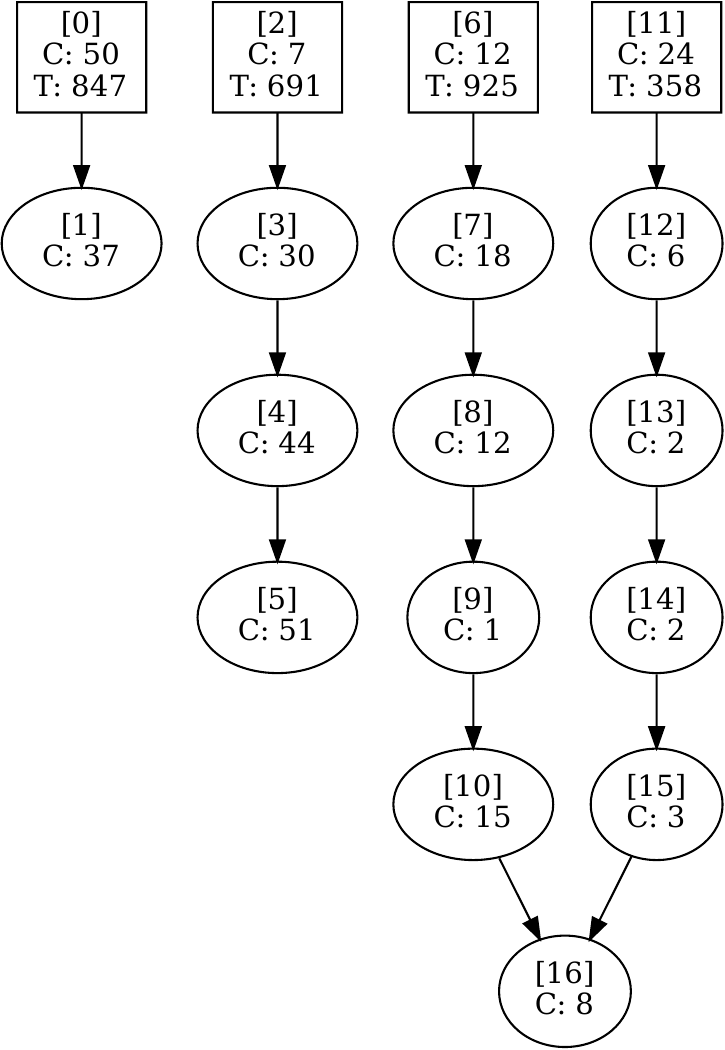}}
            \hspace{3.0cm} Number of chains: 4,\\ Total utilization: 0.5
        \end{minipage}
        % 3
        \begin{minipage}{0.36\linewidth}
            \centering
            \scalebox{1.0}{\includegraphics[keepaspectratio, width = \linewidth]{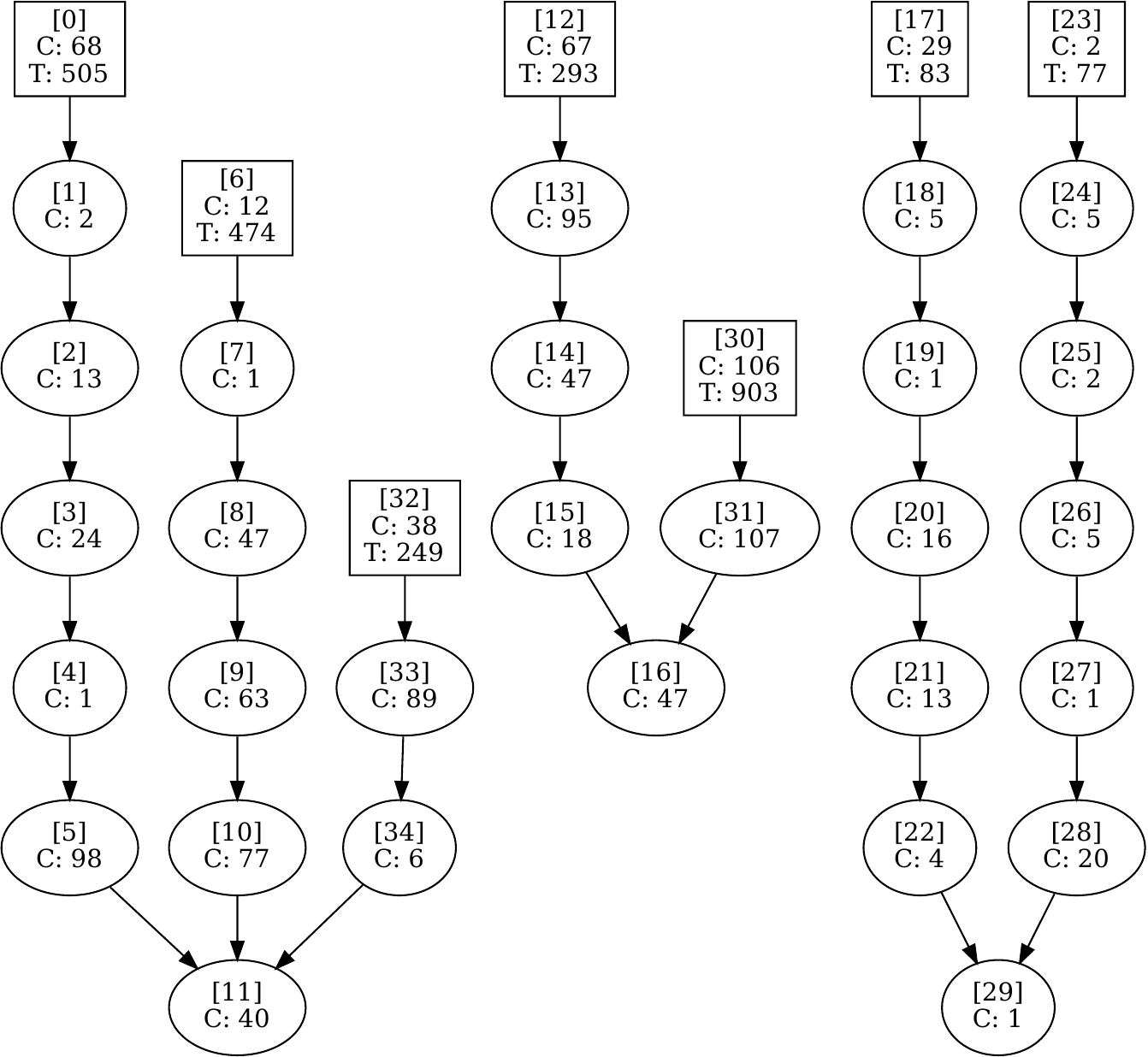}}
            \hspace{3.5mm} Number of chains: 7,\\ Total utilization: 4.0
        \end{minipage}
    \end{tabular}
    \centering
    \vspace{-1mm}
    \caption{Example of generating chain-based DAGs of various total utilization.}
    \label{fig: case_study_chain}
    \vspace{-6mm}
\end{figure*}

\subsubsection{Case Study 2: Multi-rate DAGs with Various Total Utilization}
\label{ssec: case_study_2}

Random DAG sets with different total utilization are used in the evaluation of most studies considering multi-rate DAGs.
An example of generating a multi-rate DAG with a random period and the execution time for each node based on the total utilization of the DAG, as used in existing studies~\cite{gunzel2021suspension, ueter2021hard}, is shown in Fig.~\ref{fig: case_study_all_timer}.
Here, timer-driven nodes are drawn as squares.
Because {\it Combination} is specified for {\it Total utilization} in Fig.~\ref{fig: case_study_all_timer}, RD-Gen generates \rr{1.2}{100 DAGs for each 5\% increment in total utilization from 5\% to 95\%.}

For each DAG, the utilization of each node is determined by the UUniFast method to satisfy the total utilization, and the execution time is calculated from the utilization and a randomly selected period from a uniform distribution ranging from 1 to 100.

\subsubsection{Case Study 3: Chain-based DAGs with Various Total Utilization}
\label{ssec: case_study_3}

In the evaluation of chain-based DAGs, a random DAG set is used with varying total utilization for the entire system and each chain.
An example of generating a random DAG set, as used in existing studies~\cite{tang2020response, choi2021picas}, in which the utilization of each chain is determined from the total utilization of the entire system, and the chain period and the execution time of each node is randomly assigned to satisfy the utilization of the chain is shown in Fig.~\ref{fig: case_study_chain}.
RD-Gen randomly sets the utilization of each chain using the UUniFast method to achieve a determined total utilization.
Here, since {\it Maximum utilization} is specified as 1.0 in Fig.~\ref{fig: case_study_chain}, RD-Gen never generates a chain with utilization greater than 1.0.
The period of each chain is randomly set in the range of 50 to 1${,}$000, and the execution time of each node is calculated based on the utilization and period (lines 10--18 in Algorithm~\ref{alg: set_utilization}).

\vspace{-1mm}
\section{Evaluation}
\label{sec: evaluation}
\vspace{-1mm}

This section shows that compared to existing random DAG generation tools, RD-Gen can generate a random DAG set with fewer lines of code and no unique user implementation.
Table~\ref{tab: processes} shows the processes that must be implemented by the user and the number of lines of code in our implementation.
Here, the process in Table~\ref{tab: processes} is written in Python and shell scripts, and the Python NetworkX library is used for processes related to DAGs.

The amount of description required to generate a random DAG set for the case studies in Section~\ref{ssec: case_study_1} is shown in Table~\ref{tab: eval_case_study_single}.
In this case, while TGFF and GGen require the user to implement more than 50 lines, RD-Gen can generate all random DAG sets using only the tool's functionality.
Since TGFF and GGen do not have the functionality to generate DAGs with different parameter settings at once, the user must execute the generation command many times while changing the parameters, using shell scripts or other means.
In contrast, RD-Gen generates DAGs with all parameter combinations by specifying {\it Combination} as a parameter and automatically divides them into directories.
Although TGFF uses the {\it Fan-in/Fan-out} method to construct DAGs, the .tgff files output by TGFF are in a proprietary format and require users to implement a reading process.
Furthermore, since TGFF and GGen do not guarantee that the graph is weakly connected, users may unintentionally use a non-weakly connected DAG for evaluation.
TGFF and GGen cannot specify parameters such as CCR that are calculated from multiple properties. Therefore, the CCR must be adjusted for the DAG set after it is generated.
In RD-Gen, users can automate all of the above processes by inputting the YAML file as shown on the left side of Fig.~\ref{fig: case_study_single}.

\begin{table}[tb]
    \centering
    \caption{Processes that must be implemented by the user \\ and lines of code}
    \label{tab: processes}
    \vspace{-3mm}
    \renewcommand{\arraystretch}{1.0}
    \scalebox{0.8}{
        \begin{tabular}{|c|l|c|}\hline
            \textbf{Symbols} & \MC{1}{c|}{\textbf{Processes}}                                & \textbf{Lines of code} \\\hline
            $(a)$            & Loop with different parameter values                          & 3                      \\\hline
            $(b)$            & Separate DAGs generated for each parameter into directories   & 5                      \\\hline
            $(c)$            & Load .tgff file as DAG                                        & 40                     \\\hline
            $(d)$            & Construct a DAG using G(n, p) method                          & 10                     \\\hline
            $(e)$            & Construct a DAG using chain-based method                      & 80                     \\\hline
            $(f)$            & Merge the sink nodes to a specific number                     & 10                     \\\hline
            $(g)$            & Add edges to be weakly connected                              & 15                     \\\hline
            $(h)$            & Set the properties randomly within a specific range           & 10                     \\\hline
            $(i)$            & \tabml{Set the execution time for each node and                                        \\ the communication time for each edge based on the CCR}                             & 25            \\\hline
            $(j)$            & \tabml{Set the utilization for a multi-rate DAG consisting of                          \\ only timer-driven nodes (lines 1--7 in Algorithm~\ref{alg: set_utilization})}                                             & 30            \\\hline
            $(k)$            & \tabml{Set the utilization for a chain-based DAG                                       \\ (lines 8--19 in Algorithm~\ref{alg: set_utilization})}                                             & 50            \\\hline
        \end{tabular}
    }
    \vspace{0mm}
\end{table}

\begin{table}[tb]
    \centering
    \caption{Effort required to generate random DAG sets for {\it Case Study 1}~\cite{subbaraj2020multi, liu2016minimizing, sheikh2016sixteen}}
    \label{tab: eval_case_study_single}
    \vspace{-3mm}
    \renewcommand{\arraystretch}{1.0}
    \scalebox{0.85}{
        \begin{tabular}{|c|c|l|}
            \hline
            \multicolumn{1}{|l|}{}         & \multicolumn{1}{c|}{\begin{tabular}[c]{@{}c@{}}\textbf{Lines of description}\\ \textbf{for tool options}\end{tabular}} & \multicolumn{1}{c|}{\textbf{Lines of code}}         \\ \hline
            TGFF~\cite{tgff}               & 6                                                                                                                      & \tabml{$(a) \times 2 + (b) + (c) + (f) + (g) + (i)$ \\ $= 101$} \\ \hline
            GGen~\cite{cordeiro2010random} & 3                                                                                                                      & \tabml{$(a) \times 2 + (b) + (g) + (i)$             \\ $= 51$}              \\ \hline
            RD-Gen                         & 20                                                                                                                     & 0                                                   \\ \hline
        \end{tabular}
    }
    \vspace{1mm}
\end{table}

\begin{table}[tb]
    \centering
    \caption{Effort required to generate random DAG sets for {\it Case Study 2}~\cite{gunzel2021suspension, ueter2021hard}}
    \label{tab: eval_case_study_all_timer}
    \vspace{-3mm}
    \renewcommand{\arraystretch}{1.0}
    \scalebox{0.85}{
        \begin{tabular}{|c|c|l|}
            \hline
            \multicolumn{1}{|l|}{}         & \multicolumn{1}{c|}{\begin{tabular}[c]{@{}c@{}}\textbf{Lines of description}\\ \textbf{for tool options}\end{tabular}} & \multicolumn{1}{c|}{\textbf{Lines of code}} \\ \hline
            TGFF~\cite{tgff}               & -                                                                                                                      & \tabml{$(a) + (b) + (d) + (g) + (j)$        \\ $= 63$} \\ \hline
            GGen~\cite{cordeiro2010random} & 2                                                                                                                      & \tabml{$(a) + (b) + (g) + (j)$              \\ $= 53$}       \\ \hline
            RD-Gen                         & 20                                                                                                                     & 0                                           \\ \hline
        \end{tabular}
    }
    \vspace{0mm}
\end{table}

The effort required to generate a random DAG set for the case studies in Section~\ref{ssec: case_study_2} is shown in Table~\ref{tab: eval_case_study_all_timer}.
TGFF does not support the {\it G(n, p)} method and multi-rate DAG.
GGen allows properties to be randomly set to nodes or edges within a user-specified range after graph construction.
However, GGen does not allow users to specify the total utilization used in the evaluation of most studies that consider multi-rate DAGs.
Therefore, the user has to go through the trouble of implementing such as lines 1--7 in Algorithm~\ref{alg: set_utilization}.
In contrast, RD-Gen uses the UUniFast method to randomly and automatically set the period and the execution time for each node to meet the specified total utilization.

\begin{table}[tb]
    \centering
    \caption{Effort required to generate random DAG sets for {\it Case Study 3}~\cite{tang2020response, choi2021picas}}
    \label{tab: eval_case_study_chain}
    \vspace{-3mm}
    \renewcommand{\arraystretch}{1.0}
    \scalebox{0.85}{
        \begin{tabular}{|c|c|l|}
            \hline
            \multicolumn{1}{|l|}{}         & \multicolumn{1}{c|}{\begin{tabular}[c]{@{}c@{}}\textbf{Lines of description}\\ \textbf{for tool options}\end{tabular}} & \multicolumn{1}{c|}{\textbf{Lines of code}} \\ \hline
            TGFF~\cite{tgff}               & -                                                                                                                      & \tabml{$(a) + (b) + (d) + (k)$              \\ $= 138$}      \\ \hline
            GGen~\cite{cordeiro2010random} & -                                                                                                                      & \tabml{$(a) + (b) + (d) + (k)$              \\ $= 138$}      \\ \hline
            RD-Gen                         & 22                                                                                                                     & 0                                           \\ \hline
        \end{tabular}
    }
    \vspace{1mm}
\end{table}

The amount of description to generate a chain-based random DAG set, as in the case study in Section~\ref{ssec: case_study_3}, is shown in Table~\ref{tab: eval_case_study_chain}.
\rr{1.6}{Chain-based DAGs, which are considered in research on state-of-the-art self-driving systems, cannot be generated by the classic random DAG generation tools, TGFF and GGen.}
RD-Gen can generate a batch of random chain-based DAG sets with different total utilization without any user implementation by inputting a YAML file as shown in Fig.~\ref{fig: case_study_chain}.

The evaluation results demonstrated that RD-Gen could generate the random DAG sets used in the evaluation with only an intuitive YAML file description, regardless of the single-rate and multi-rate cases.
Therefore, RD-Gen is a flexible evaluation platform that can be adapted to various requirements and provides reliability and reproducibility for the latest DAG studies.

\vspace{-3mm}
\section{Related work}
\label{sec: related_work}
\vspace{-1mm}

This section describes existing random DAG generation tools and existing studies using random DAGs and compares them with RD-Gen.
Table~\ref{tab: comparison_rd_gen} shows a comparison of RD-Gen with existing methods.

\begin{table}[tb]
    \centering
    {
        \caption{RD-Gen VS. existing methods}
        \label{tab: comparison_rd_gen}
        \vspace{-3mm}
        \renewcommand{\arraystretch}{1.0}
        \scalebox{0.8}{
            \begin{tabular}{|c|c|c|c|c|c|c|} \hline
                                                           & \textbf{RSD} & \textbf{RMD} & \textbf{RCD} & \textbf{RPU} & \textbf{RDT} & \textbf{OCG} \\\hline
                TGFF~\cite{tgff}                           & \ch          &              &              &              & \ch          &              \\\hline
                GGen~\cite{cordeiro2010random}             & \ch          & \ch          &              &              & \ch          &              \\\hline
                DAGEN~\cite{amalarethinam2011dagen}        & \ch          &              &              &              & \ch          &              \\\hline
                MRTG~\cite{ashish2016modular}              & \ch          &              &              &              & \ch          &              \\\hline
                Voronov et al.~\cite{voronov2021ai}        &              & \ch          &              & \ch          &              &              \\\hline
                Dong et al.~\cite{dong2019efficient}       &              & \ch          &              & \ch          &              &              \\\hline
                He et al.~\cite{he2021response}            & \ch          &              &              & \ch          &              &              \\\hline
                Yang et al.  ~\cite{yang2020mixed}         &              & \ch          &              & \ch          &              &              \\\hline
                Gunzel et al.~\cite{gunzel2021suspension}  &              & \ch          &              & \ch          &              &              \\\hline
                Ueter et al.~\cite{ueter2021hard}          &              & \ch          &              & \ch          &              &              \\\hline
                Verucchi et al.~\cite{verucchi2020latency} &              & \ch          &              & \ch          &              &              \\\hline
                Klaus et al.~\cite{klaus2021constrained}   &              & \ch          &              & \ch          &              &              \\\hline
                Kordon et al.~\cite{kordon2020evaluation}  &              & \ch          &              &              &              &              \\\hline
                Tang et al.~\cite{tang2020response}        &              & \ch          & \ch          & \ch          &              &              \\\hline
                Choi et al.~\cite{choi2021picas}           &              & \ch          & \ch          & \ch          &              &              \\\hline
                RD-Gen                                     & \ch          & \ch          & \ch          & \ch          & \ch          & \ch          \\\hline
            \end{tabular}
        }
        \begin{tablenotes}[normal]{
                \scriptsize
                \item {RSD}: Random generation of single-rate DAGs
                \vspace{-0mm}
                \item {RMD}: Random generation of multi-rate DAGs
                \vspace{-0mm}
                \item {RCD}: Random generation of chain-based DAGs
                \vspace{-0mm}
                \item {RPU}: Random property settings based on total utilization
                \vspace{-0mm}
                \item {RDT}: Random DAG generation tool
                \vspace{-0mm}
                \item {OCG}: One-command batch generation of random DAGs
            }
        \end{tablenotes}
    }
    \vspace{1mm}
\end{table}

\vspace{-1mm}
\subsection{Random DAG Generation Tools}
\label{sec: random_tool}
\vspace{-1mm}

Random DAG generation tools provide reliability and reproducibility for evaluations of scheduling and latency analysis studies.
TGFF~\cite{tgff} is the first tool proposed for this purpose and has been used to evaluate recent studies~\cite{roeder2021energy, fard2021analytical, wu2021evolutionary}.
TGFF determines the shape of a DAG primarily by specifying either or both the maximum and minimum input degree and maximum (first) output degree for a single node (the {\it Fan-in/Fan-out} method).
TGFF can quickly generate many DAGs, and the task set can be easily reproduced by other researchers by inputting the same parameters.
However, TGFF was released in 1998 and has many problems, such as its output format (.tgff), which is difficult to handle, and it cannot generate multi-rate DAGs.

GGen~\cite{cordeiro2010random} is a unified implementation of classical task graph generation methods used in the scheduling domain.
GGen allows the user to add properties such as the period and the communication time to nodes and edges after generating DAGs using a user-specified generation method.
However, GGen does not allow constraints to be specified between different properties, such as implicit deadlines (i.e., the execution time of a node must not exceed its period).
Therefore, users with such requirements must adjust these values themselves.

Other random DAG generation tools such as DAGEN~\cite{amalarethinam2011dagen} and MRTG~\cite{ashish2016modular} have also been proposed.
DAGEN generates random workflow applications by specifying the node load balancing, edge connection probability, and workflow shape.
MRTG is a random DAG generation tool with a module-based implementation for user extensibility.
However, these tools are not capable of generating multi-rate DAGs.
Conversely, RD-Gen can flexibly generate multi-rate DAGs of various types.
In addition, RD-Gen can automatically set properties calculated by multiple values such as CCR and the total utilization.

\vspace{-1mm}
\subsection{Unique Implementation of Random DAG Generation}
\vspace{-1mm}

This section presents existing studies in which the authors generated random DAG sets to evaluate their own implemented algorithms and settings.
Because there are no random DAG generation tools that can generate multi-rate DAGs by specifying the total utilization, as described in Section~\ref{sec: random_tool}, most studies that consider multi-rate DAGs include unique implementations.

In real-time systems such as in-vehicle systems and self-driving systems, multi-rate DAGs consisting of only timer-driven nodes are considered.
Many real-time system researchers use the {\it G(n, p)} method to construct DAGs and generate random task sets with different numbers of nodes, different total utilizations, and different periods~\cite{voronov2021ai, he2021response, dong2019efficient}.
While proprietary algorithms may be used to construct graphs, the approach is similar in that the utilization is determined using the UUniFast method, and the WCET value of the nodes is assigned based on the utilization and the periods~\cite{yang2020mixed, gunzel2021suspension, ueter2021hard}.

In a multi-rate DAG, as considered in automotive systems, a period is randomly assigned to each node based on the period observed in the automotive application.
Verucchi et al.~\cite{verucchi2020latency} randomly extended the automotive benchmark proposed by BOSCH in the 2015 WATERS Challenge~\cite{kramer2015real} to analyze the performance of their proposed method.
Verucchi et al. randomly set the task period from the values \rr{2.6}{$[1, 5, 10, 20, 50, 100, 200, 1000]$} in milliseconds, as found in automotive applications, for DAG utilization.
Klaus et al.~\cite{klaus2021constrained} set the utilization, the period, and the number of nodes for each DAG node.
Verucchi et al. and Klaus et al. create random task sets based on node chains consisting of only timer-driven nodes.
RD-Gen can also generate chains consisting of only timer-driven nodes using the {\it Chain-based} method and by specifying the {\it Periodic type} as {\it \dq{All.}}
Using the Python NetworkX library, Kordon et al.~\cite{kordon2020evaluation} randomly set the period, the number of edges per task, the release time, and the number of source nodes for their DAG sets.
RD-Gen can automatically set all combinations of the total utilization, periods, and the number of nodes as described above.

Studies of chain-based DAGs, such as the latest ROS-based systems, have also been evaluated using random task sets.
Tang et al.~\cite{tang2020response} allocated a value of the utilization to each chain based on the total utilization of the entire system and the number of chains and then assigned the utilization to the execution units in the chain.
Choi et al.~\cite{choi2021picas} similarly assigned a utilization value to each chain using the UUniFast method based on the total system-wide utilization.
Multi-rate DAGs based on such chains can be generated flexibly with the {\it Chain-based} method in RD-Gen.
In addition, because RD-Gen allows the utilization to be automatically set based on the chain, researchers do not need to implement this functionality on their own.

\vspace{-1mm}
\section{Conclusions}
\label{sec: conclusion}
\vspace{-1mm}

In this paper, we have proposed a random DAG generator considering multi-rate applications for reproducible scheduling evaluation called RD-Gen that can generate both single-rate DAGs and state-of-the-art multi-rate DAGs.
RD-Gen extended the existing random graph generation methods for DAGs and provided a new chain-based method.
RD-Gen could automatically set complex properties such as CCR and the total utilization.
% Moreover, RD-Gen could support researchers by providing functions such as the batch generation of a given number of DAG sets for different parameters.
Case studies indicated that RD-Gen could meet the requirements of DAG studies in a variety of problem settings.
% Therefore, RD-Gen can provide reliability and easy reproducibility for DAG studies.
In future work, we plan to extend RD-Gen to cover additional graph generation methods and more complex properties.

%\section*{Acknowledgment}
%\todo{}

\renewcommand{\baselinestretch}{1.0}
\bibliography{./bibliography/master_reference}

\end{document}